\documentclass[11pt]{article}
\pdfoutput=1
\usepackage{soul}
\usepackage{jheppub}
\usepackage{amsfonts}
\usepackage{amsmath}
\usepackage{amsthm}
\allowdisplaybreaks[4]         
\usepackage{amssymb}   
\usepackage{euscript}   
\usepackage{cleveref}    
\usepackage[dvipsnames]{xcolor}          
\usepackage{tensor}     
\usepackage{graphicx}
\usepackage{caption}
\usepackage{subcaption}
\usepackage[skip=0.5ex, belowskip=1ex,
                labelformat=brace,
                singlelinecheck=off]{subcaption}   
\usepackage{float}
\usepackage{hyperref}
\hypersetup{
	colorlinks =NavyBlue,
	linkcolor =NavyBlue,
	citecolor=MidnightBlue,
	filecolor=NavyBlue,
	urlcolor=NavyBlue,
}

\newcommand{\bea}{\begin{eqnarray}}
\newcommand{\eea}{\end{eqnarray}}
\newcommand{\ba}{\begin{eqnarray}}
\usepackage{braket}
\newcommand{\ea}{\end{eqnarray}}

\newcommand{\beq}{\begin{equation}}
\newcommand{\eeq}{\end{equation} }
\newcommand{\beqa}{\begin{eqnarray}}

\newcommand{\eeqa}{\end{eqnarray}}
\newcommand{\beqar}{\begin{eqnarray*}}
\newcommand{\eeqar}{\end{eqnarray*}}

\newcommand{\be}{\begin{equation}}
\newcommand{\ee}{\end{equation}}
\newcommand{\diff}{\mathrm{d}}

\newcommand{\Lag}{\mathcal{L}}

\definecolor{shadecolor}{rgb}{.25,.25,.25}

\preprint{\texttt{WI-27-2024, IFT-UAM/CSIC-24-123}}

\title{ \boldmath Kasner eons with matter: holographic excursions to the black hole singularity}

\author[a]{Elena C\'aceres,}
\author[b]{\'Angel J. Murcia,}
\author[c]{Ayan K. Patra,}
\author[c]{Juan F. Pedraza}

\affiliation[a]{Theory Group, Department of Physics, University of Texas, Austin, TX 78712, USA
 \vspace{0.1cm}}
\affiliation[b]{INFN, Sezione di Padova, Via Francesco Marzolo 8, I-35131 Padova, Italy \vspace{0.1cm}}
\affiliation[c]{Instituto de Física Teórica UAM/CSIC, Calle Nicol\'as Cabrera 13-15, Madrid, E-28049, Spain}

\emailAdd{elenac@utexas.edu}
\emailAdd{angel.murcia@pd.infn.it}
\emailAdd{a.patra@csic.es}
\emailAdd{j.pedraza@csic.es}

\abstract{Recent work has shown that introducing higher-curvature terms to the Einstein-Hilbert action causes the approach to a space-like singularity to unfold as a sequence of \emph{Kasner eons}. Each eon is dominated by emergent physics at an energy scale controlled by higher-curvature terms of a given order, transitioning to higher-order eons as the singularity is approached. The purpose of this paper is twofold. First, we demonstrate that the inclusion of matter dramatically modifies the physics of eons compared to the vacuum case. We illustrate this by considering a family of quasi-topological gravities of arbitrary order minimally coupled to a scalar field. Second, we investigate Kasner eons in the interior of black holes with field theory duals and analyze their imprints on holographic observables. We show that the behavior of the thermal $a$-function, two-point functions of heavy operators, and holographic complexity can capture distinct signatures of the eons, making them promising tools for diagnosing stringy effects near black hole singularities.}

\begin{document} 
\maketitle
\flushbottom

\section{Introduction}

General relativity has long been known to predict a complex but manageable behavior near space-like singularities, particularly in the Belinski-Khalatnikov-Lifshitz (BKL) limit \cite{Belinsky:1970ew,Belinskii:1973zz,Belinskii:1981vdw}, where spatial derivatives become negligible and ultra-local `Carrollian' dynamics emerge \cite{Oling:2024vmq}. The Kasner metric, a key solution in this limit, describes a homogeneous yet anisotropically expanding spacetime with fixed scaling exponents, constrained by specific relations to satisfy Einstein's equations. Under certain simplifying assumptions, this metric is conjectured to characterize the `final regime' of evolution before general relativity loses its predictive power.

More generally, the BKL limit predicts a sequence of Kasner epochs ---periods where the spacetime metric closely follows a Kasner geometry--- connected by brief transitions. The dynamics of the BKL limit are particularly intriguing when the scaling exponents evolve through a space of solutions defined by the Kasner relations. This evolution can be influenced by factors like spatial curvature or couplings to specific matter fields. For example, in the context of supergravity, it has been shown that the inclusion of $p$-forms triggers `mixmaster' dynamics governed by affine Lie algebras~\cite{Damour:2002et}. In this scenario, the evolution of Kasner exponents can be understood as a particle moving on a hyperbolic billiard table, giving rise to the concept of `cosmological billiards.' The boundaries of this table, which define the symmetry algebra, are determined by the particular setup under consideration. 

The long-term evolution of the BKL limit reveals two key stages: Kasner eras and Kasner eons. A Kasner era comprises multiple Kasner epochs, where two spatial directions alternately contract and expand while the others remain fixed \cite{Belinski:2017fas}. However, numerical evolution through these eras is not so straightforward, due to the emergence of `spikes,' spatial structures that disrupt expected asymptotic locality \cite{Berger:2002st,Heinzle:2012um,Czuchry:2016rlo}. Recently, higher-curvature `stringy' corrections to the Einstein-Hilbert action have been shown to introduce new transitions, causing the approach to a space-like singularity to unfold through a sequence of Kasner eons ---extended periods dominated by physics at energy scales defined by the higher-curvature terms \cite{Bueno:2024fzg}. As the singularity nears, the evolution progresses through eons of increasingly higher order.

The analysis in \cite{Bueno:2024fzg} focused on specific higher-order gravities in vacuum. While the emergence of eons is likely a robust feature arising from emergent stringy effects, it is essential to recognize the critical role that matter fields play in any BKL analysis. As space-like singularities are approached, matter fields may undergo infinite growth, resulting in substantial backreaction on the geometry. This instability typically modifies the Kasner relations, which can vary significantly depending on the matter content of the theory~\cite{Belinsky:1970ew,Belinskii:1973zz,Belinskii:1981vdw}. Therefore, incorporating matter is crucial for understanding the dynamics near space-like singularities, and should be essential for fully uncovering the physics of Kasner eons. To address this issue, in the first part of this paper we will investigate a family of quasi-topological gravities of arbitrary order coupled to a scalar field, illustrating how higher-order gravitational effects interact with matter fields and influence behavior near singularities.\footnote{In pure GR, this model only leads to a unique Kasner regime at late times. More generally, it would be interesting to consider matter content that gives rise to `mixmaster' behavior. This would allow for a deeper understanding of the interplay between Kasner epochs, eras, and eons. We leave this study for future work.} These findings will offer valuable insights into Kasner eons from a purely gravitational perspective.

In the second part of this paper, we shift our focus to Kasner eons within the context of the AdS/CFT correspondence. It is well-known that the singularity inside an AdS-Schwarzschild black hole exhibits Kasner-like behavior. Recently, Frenkel and collaborators \cite{Frenkel:2020ysx} studied a class of AdS black holes with scalar hair, uncovering a deformation of the singularity into a more general Kasner universe, consistent with BKL expectations. Following \cite{Frenkel:2020ysx}, several studies have studied the dynamics of Kasner singularities from a holographic perspective, including scenarios with cosmological bounces and Kasner inversions akin to those observed in mixmaster universes \cite{Hartnoll:2020rwq,Hartnoll:2020fhc,Caceres:2021fuw,Bhattacharya:2021nqj,Sword:2021pfm,Sword:2022oyg,Wang:2020nkd,Mansoori:2021wxf,Liu:2021hap,Das:2021vjf,Caceres:2022smh,An:2022lvo,Auzzi:2022bfd,Mirjalali:2022wrg,Hartnoll:2022snh,Hartnoll:2022rdv,Caceres:2022hei,Liu:2022rsy,Gao:2023zbd,Caceres:2023zhl,Blacker:2023ezy,Caceres:2023zft,DeClerck:2023fax,Cai:2023igv,Gao:2023rqc,Arean:2024pzo,Cai:2024ltu,Carballo:2024hem}. Within the AdS/CFT correspondence, these phenomena are interpreted in terms of Renormalization Group (RG) flows of the dual CFT \cite{Caceres:2022smh}, where various holographic observables capture essential information about the near-singularity geometries. Based on these antecedents, we thus anticipate that we can engineer specific observables to identify the presence of eons from the field theory perspective.

This paper is organized as follows. In section \ref{sec:sec2} we revisit the concept of Kasner eons, illustrating their emergence in a specific family of quasi-topological gravities that include terms of arbitrary order in curvature and an arbitrary cosmological constant.  We investigate Kasner eons in the vacuum of these theories and in the presence of a scalar field, where, as anticipated, we uncover important modifications of the relations that the Kasner exponents must satisfy. Notably, when matter is included, these theories ensure that any static and plane-symmetric background is fully described by second-order equations of motion, which drastically simplifies our analysis. In section \ref{sec:hol} we focus on Kasner eons within black holes with a negative cosmological constant and AdS boundary conditions, which are conducive to a holographic interpretation. We analyze a handful of observables designed to capture distinct signatures of the eons in terms of the dual field theory, including the thermal $a$-function, two-point functions of heavy operators, and holographic complexity. Finally, we conclude in section \ref{sec:discussion} with a summary of our findings and potential directions for future research.

\section{Kasner eons in quasi-topological gravities\label{sec:sec2}}

In their seminal work in the '70s, Belinski, Khalatnikov and Lifshitz \cite{Belinsky:1970ew,Belinskii:1973zz,Belinskii:1981vdw} showed that the classical approach to a space-like singularity ---whether cosmological or within a black hole---unfolds as a sequence of Kasner epochs, during which the metric takes the form:
\begin{equation}
    \diff s^2=-\mathrm{d}\tau^2+\sum_{i=1}^d \tau^{2p_i} \mathrm{d}x_i^2\,,
    \label{eq:kasner}
\end{equation}
where we consider a spacetime with $D = d+1$ dimensions, and the Kasner exponents $p_i$ satisfy the following relations:
\begin{equation}
    \sum_{i=1}^d p_i=1\,, \qquad \sum_{i=1}^d p_i^2=1\,.
    \label{eq:kasnergr}
\end{equation}
These conditions are needed to fulfill the Einstein equations in vacuum, but are generically altered with the inclusion of matter or higher-curvature terms. Therefore, we will refer to any spacetime of the form \eqref{eq:kasner} as Kasner, regardless of whether the Kasner exponents satisfy \eqref{eq:kasnergr}.  Kasner eras, on the other hand, consist of a sequence of Kasner epochs, during which two spatial directions successively alternate between contracting and expanding.

Meanwhile, Kasner eons emerge over longer time scales following the introduction of higher-curvature terms in the gravitational action \cite{Bueno:2024fzg}. In what follows, we will define Kasner eons and demonstrate how a potentially infinite sequence naturally arises as the gravitational action is supplemented with a tower of higher-curvature corrections. First, in subsection \ref{subsec:vac}  we will focus on purely gravitational theories without matter, extending the pioneering work of \cite{Bueno:2024fzg} to arbitrary derivative orders and spacetime dimensions $D \geq 5$. Next, in subsection \ref{subsec:mat}, we will introduce a minimally coupled scalar field and explore how its presence modifies the approach to the singularity by examining the Kasner exponents across different eons.

The computations and derivations in this section are conducted from a purely gravitational perspective. Specifically, Kasner eons are identified by analyzing the bulk equations of motion within the higher-curvature theories under consideration. Later, in section \ref{sec:hol}, we will consider Kasner eons in the presence of a negative cosmological constant and demonstrate how they can also be probed using CFT observables via the AdS/CFT correspondence.

\subsection{Kasner eons in vacuum}
\label{subsec:vac}

Let us start by briefly explaining and motivating the emergence of Kasner eons. For now, we will focus on a vacuum scenario and consider a specific theory of gravity in $D=d+1$ dimensions that may include an infinite tower of higher-curvature corrections:
\begin{equation}
    \Lag=\frac{\lambda_0}{\ell ^2}+\lambda_1 R+\sum_{n=2}^\infty \lambda_n \ell^{2(n-1)} \mathcal{Z}_{(n)}\,.
    \label{eq:laggen}
\end{equation}
Here, $\ell$ denotes a length scale at which higher-curvature corrections become important, while $\lambda_n$ are dimensionless constants, with $\lambda_0$ representing the cosmological constant expressed in units of $\ell$.\footnote{Note that the cosmological constant decouples in regions of high curvature and is thus irrelevant to the physics of BKL singularities.} Without loss of generality, we will set $\lambda_1=1$ unless specified, achieved by rescaling the theory so that $\lambda_n \to \lambda_n/\lambda_1$. The terms $\mathcal{Z}_{(n)}$ refer to specific higher-curvature terms chosen such that the gravitational equations of motion for \emph{any} static and plane-symmetric background (equivalently, spherically symmetric or hyperbolic symmetric) in vacuum remain first-order in derivatives. These theories constitute a special subset of \emph{quasi-topological gravities} \cite{Oliva:2010eb,Myers:2010ru,Dehghani:2011vu,Bueno:2019ltp,Bueno:2019ycr,Moreno:2023rfl,Moreno:2023arp},\footnote{Quasi-topological gravities are defined so that for single-function static and plane-symmetric solutions ---such as the Schwarzschild-Tangherlini solution---, the equation of motion for the unique function determining the metric can always be integrated into an algebraic equation. Requiring that the equations of motion for \emph{any} static and plane-symmetric background in vacuum are first-order in derivatives imposes a stronger condition.} representing the most natural generalization of Lovelock gravities for curvature orders $2n > D$. Interestingly enough, one can prove (following the arguments from \cite{Bueno:2019ltp}, although this reference proves something slightly different) that any effective theory of gravity with no derivatives of the curvature may be mapped, order by order in the UV scale $\ell$, to a theory of the form given in \eqref{eq:laggen}. As discussed in \cite{Bueno:2024fzg}, examples of these theories ---with first-order equations for any static and plane-symmetric background in vacuum--- exist at all curvature orders and for dimensions $D \geq 5$. For the reader's convenience, we provide in Appendix \ref{app:quasith} specific expressions for $\mathcal{Z}_{(n)}$ up to fifth order in the curvature for any spacetime dimension $D \geq 5$ that satisfies our criteria, and discuss how to construct $\mathcal{Z}_{(n)}$ for $n > 5$.

The equations of motion for the theory \eqref{eq:laggen} can always be expressed in the following compact form ---see, e.g., \cite{Padmanabhan:2013xyr}:
\begin{equation}
    P_{acde} R_b{}^{cde}-\frac{1}{2} g_{ab} \mathcal{L}+2 \nabla^{c} \nabla^{d} P_{acbd}=0\,, \qquad P^{abcd}\equiv\frac{\partial \mathcal{L}}{\partial R_{abcd}}\,.
\end{equation}
We will adopt the following static and plane-symmetric ansatz for the spacetime metric:
\begin{equation}
\label{eq:bbans}
    \mathrm{d}s^2=\frac{1}{r^2} \left[ - e^{-\chi(r)}f(r) \diff t^2+\frac{\diff r^2}{f(r)}+ \mathrm{d} \vec{x}_{d-1}^2 \right]\,.
\end{equation}
where $\mathrm{d} \vec{x}_{d-1}^2$ denotes the $(d-1)$-dimensional Euclidean metric. In these coordinates, the singularity is located at $r \rightarrow +\infty$. A key feature of the special higher-curvature theories under consideration \eqref{eq:laggen} is that $\nabla^d P_{acbd} \big|{\chi,f}=0$, where $\big|{\chi,f}$ denotes evaluation on \eqref{eq:bbans}. As a result, the complete set of gravitational equations of motion can be exactly integrated, yielding the following two conditions\footnote{This result could also have been derived from a reduced action principle, see e.g. \cite{Bueno:2022res}.}
\begin{equation}
\label{eq:eomfqt}
    \sum_{n=0}^\infty \lambda_n  (-1)^n f(r)^n r^{-d} (d+1-2n)(\ell^2 d(d+1))^{n-1}=C\,, \qquad \chi(r)=\chi_0\,.
\end{equation}
Here $\chi_0$ is a constant that can be arbitrarily set to zero through a time reparametrization and $C$ is a constant related to the energy of the spacetime. By solving \eqref{eq:eomfqt}, an algebraic equation for $f(r)$, the exact solution for the metric can be determined, enabling a detailed analysis of the approach to the singularity ---provided it exists. However, an analytical solution for $f(r)$ is generally not possible, complicating a direct study of the singularity.

The behavior of the solutions can nevertheless be examined perturbatively as the singularity is approached. In the initial stage, far from the singularity, Einstein gravity dominates. However, as the singularity is approached, higher-curvature corrections become increasingly significant, gradually taking control of the dynamics. First, quadratic-curvature terms surpass the Einstein regime, determining the new Kasner exponents and the corresponding Kasner metric. As one gets even closer, cubic terms dominate, making the quadratic terms negligible and fully controlling the solution. Each phase where a specific curvature order dominates is called a \emph{Kasner eon} \cite{Bueno:2024fzg}, resulting in a sequence of Kasner eons transitioning from one another as the singularity is approached. From \eqref{eq:eomfqt} it follows that, sufficiently far from these transitions, the $n$-th eon ---obtained by setting all $\lambda_{m\neq n}$ to zero---  is defined by:
\begin{equation}
    f(r) \sim -r^{d/n}\,.
    \label{eq:kasnerexpfvac}
\end{equation}
Each Kasner eon can be characterized through the corresponding Ricci scalar derived from \eqref{eq:bbans}, which diverges as $r^{d/n}$ in the limit $r \rightarrow +\infty$. This indicates that every eon can be associated with a physical curvature singularity, though the severity of the singularity diminishes as $n\rightarrow +\infty$. This observation highlights the necessity of incorporating the complete tower of higher-curvature terms to potentially achieve full singularity resolution \cite{Cano:2020ezi,Bueno:2024dgm}.

Let us now derive the Kasner metric corresponding to the various Kasner eons resulting from \eqref{eq:eomfqt}. To do this, we will assume that:
\begin{equation}
\label{eq:kasneransbhvac}
    f(r)=-f_n r^{\gamma_n}\,,
\end{equation}
where $\gamma_n\equiv d/n$ and $f_n>0$. By introducing a new coordinate defined by $f_n r^{\gamma_n+2}\mathrm{d}\tau^2=\mathrm{d}r^2$, \eqref{eq:bbans} simplifies to a pure Kasner metric:
\begin{equation}
    \mathrm{d}s^2=-\mathrm{d}\tau^2+\tau^{2 p_t} \mathrm{d} t^2+ \tau^{2 p_x} \mathrm{d} x^2_{d-1}\,,\qquad p_x=\frac{2}{\gamma_n}\,, \qquad p_t=p_x-1\,.
    \label{eq:kasnergenbhvac}
\end{equation}
Thus, we conclude that the approach to the singularity is governed by a sequence of Kasner eons, characterized by the Kasner exponents satisfying:
\begin{equation}
    p_x=p_t+1=\frac{2 n}{d}\,.
    \label{eq:kasnerexpeonvac}
\end{equation}
These exponents alter the vacuum Kasner relations of GR \eqref{eq:kasnergr} as follows:
\begin{equation}\label{eq:kasnerrelnew}
    \sum_{i=1}^d p_i=2n-1\,, \qquad  \sum_{i=1}^d p_i^2=\frac{4n(n-1)}{d}+1\,.
\end{equation}
For pure Lovelock metrics\footnote{Lovelock gravities constitute a subclass of the quasi-topological gravities under consideration and, as such, are included in the current analysis.} in even dimensions, $d+1=2n$, the Kasner exponents \eqref{eq:kasnerexpeonvac} take the values $p_x=1$, $p_t=0$. Conversely, in odd dimensions, $d=2n$, they are given by $p_x=\tfrac{d-1}{d}$, $p_t=-\tfrac{1}{d}$. In both cases, the Kasner metric exactly satisfies the vacuum equations of motion --- cf. \cite{Camanho:2015yqa}, where the Kasner solutions for the highest-order non-trivial Lovelock gravity in a given dimension are studied. In fact, we note that \eqref{eq:kasnergenbhvac} represents an exact solution of the equations of motion derived from \eqref{eq:laggen} for arbitrary $n$ and $d$, provided $\lambda_{m\neq n}$ are set to zero.

More generally, we find that the approach to the singularity in our theory of gravity \eqref{eq:laggen}, which includes potentially an infinite tower of higher-curvature terms, consists of a sequence of Kasner eons \eqref{eq:kasnergenbhvac} with Kasner exponents given by \eqref{eq:kasnerexpeonvac}. Specifically, the Kasner exponents $p_t$ and $p_x$ are shifted by $2/d$ as they transition from one Kasner eon to the next one.\footnote{Note that the Kasner exponents grow without bounds as we progress through the Kasner eons associated with higher-curvature terms. This suggests that our approach is inherently perturbative.}

To identify the distinct Kasner eons encountered during the approach to the singularity, it is useful to introduce the following effective Kasner exponents \cite{Bueno:2024fzg}:
\begin{equation}
    p_x^{\rm eff}=\frac{2 f(r)}{r f'(r)}\,, \qquad  p_t^{\rm eff}=\frac{2 f(r)}{r f'(r)}-1\,.
    \label{eq:effkasnervac}
\end{equation}
Indeed, whenever $f(r)$ corresponds to a proper Kasner eon in a region near the singularity ---such that its form is given by \eqref{eq:kasneransbhvac}--- the effective exponents $p_x^{\rm eff}$ and $p_t^{\rm eff}$ are naturally described by \eqref{eq:kasnerexpeonvac}. Consequently, \eqref{eq:effkasnervac} serves as a useful indicator for the presence of Kasner eons, identifiable as intervals of the radial variable $r$ where $p_x^{\rm eff}$ and $p_t^{\rm eff}$ plateau at constant values, which can be easily spotted through a graphical analysis.

To achieve this, the explicit form of $f(r)$ must be determined by solving the algebraic equation \eqref{eq:eomfqt}. However, beyond cubic order, a closed-form solution for $f(r)$ is generally unattainable, necessitating numerical methods. Notably, we find it is numerically more stable to solve the first-order differential equation obtained by differentiating \eqref{eq:eomfqt}, which yields:
\begin{equation}
\label{eq:eomdervac}
    -\frac{\lambda_0}{\ell^2} +\sum_{n=1}^\infty \lambda_n (-1)^n (d+1-2n) (d \ell^2 f(r)(d+1))^{n-1} \left (n r f'(r) -d \, f(r) \right)=0\,.
\end{equation}
The cosmological constant term $\lambda_0$ decouples in the near-singularity limit and does not affect the analysis of Kasner eons. However, anticipating the holographic studies of the next section, we set $\lambda_0$ to be negative in our numerics, ensuring an asymptotically AdS spacetime. This choice also guarantees the existence of planar black hole solutions, with $f(r)$ representing the blackening factor. Without loss of generality, then, we may assume that $f(r)$ has one zero, representing the black hole horizon, which we set to be at some $r=r_h$. Evaluating \eqref{eq:eomdervac} at this point, we find that:
\begin{equation}
    \frac{\lambda_0}{(d-1)\ell^2}+ f'_h r_h=0\,,
\end{equation}
where $f'_h=f'(r_h)$. This is equivalent to fixing the black hole temperature $T$, since $4\pi T = -f'_h$. For each value of $r_h$ and a small $\delta > 0$, we can now proceed with the numerical integration. Starting just inside the horizon at $r = r_h + \delta$, we integrate inward towards the singularity; or starting just outside the horizon at $r = r_h - \delta$, we integrate outward towards the AdS boundary. Since $f'_h$ is fully fixed by the equations of motion, we can now approximate ---with sufficiently good accuracy--- the initial value of $f(r)$ as $f(r_h \pm \delta) \simeq \pm f'_h \delta$.  

\begin{figure}[t!]
\includegraphics[scale=0.33]{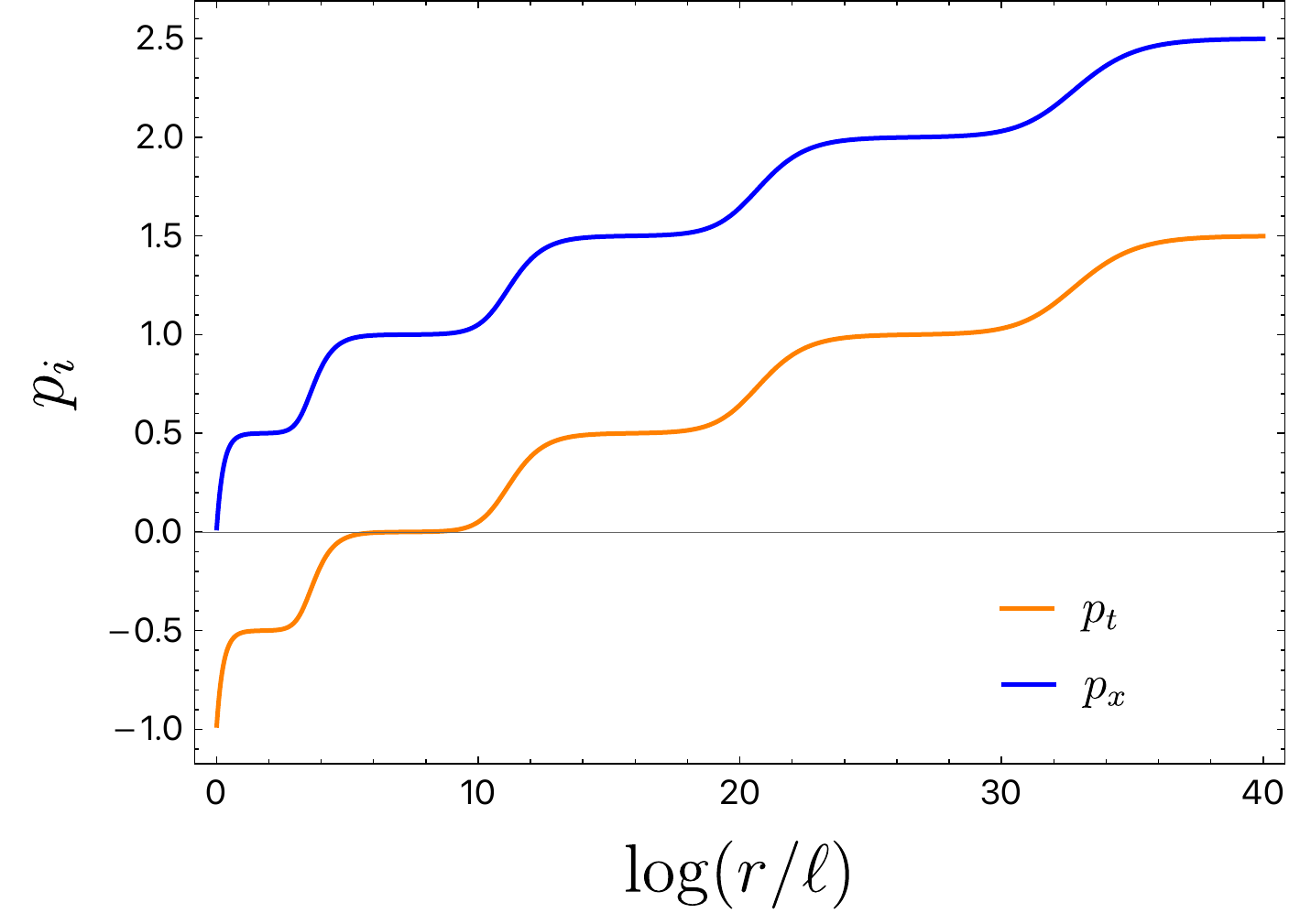}
     \includegraphics[scale=0.33]{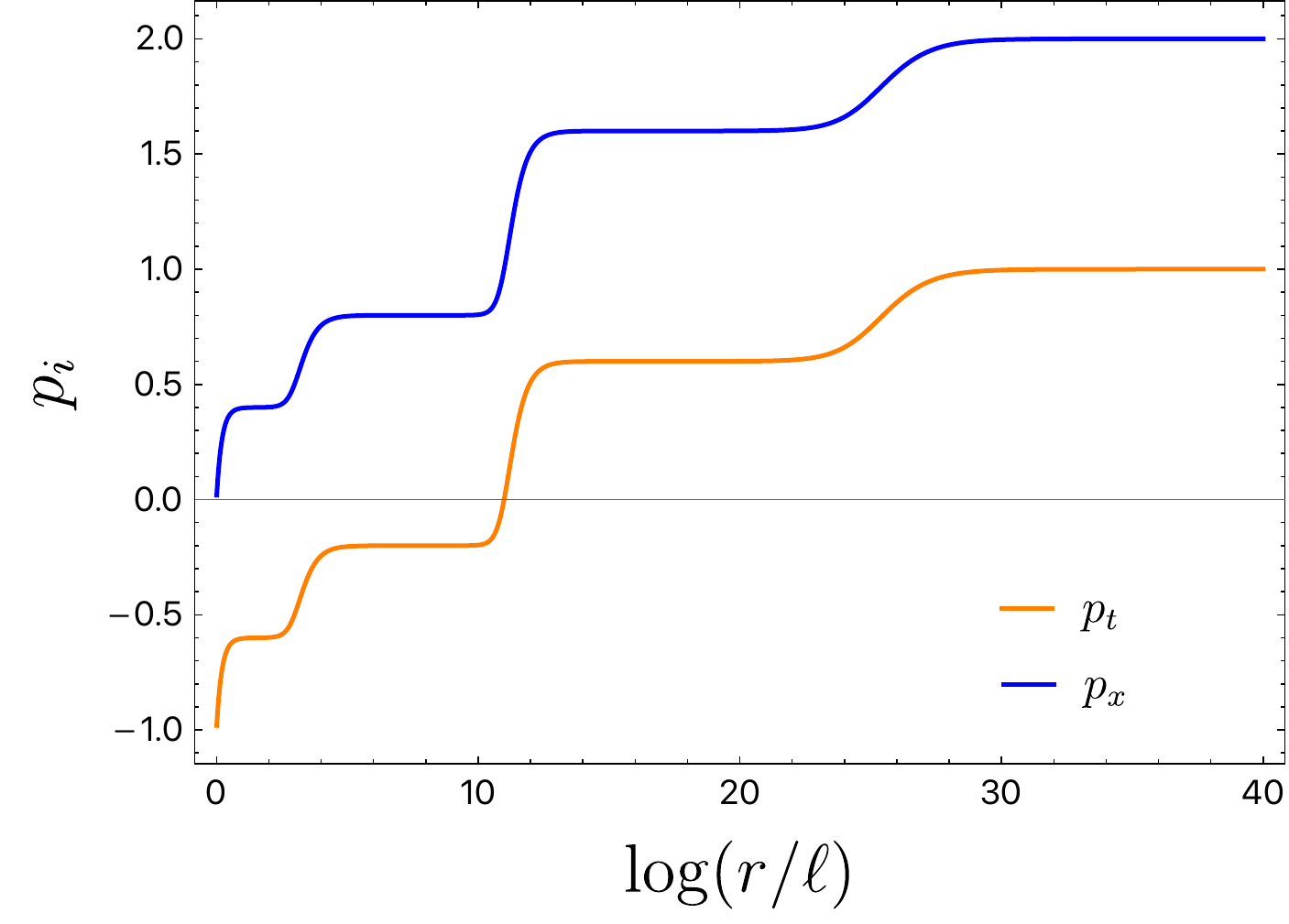}
     \caption{Kasner exponents for the higher-curvature gravities \eqref{eq:laggen} in $d=4$ (left) and $d=5$ (right) with specific dimensionless couplings $\lambda_n$ (chosen values: $\lambda_0=d(d-1)$, $\lambda_1=1$, $\lambda_2=10^{-8}$, $\lambda_3=-10^{-22}$, $\lambda_4=-10^{-42}$, $\lambda_5=-10^{-67}$ and $\lambda_{n>5}=0$). We have also set $r_h=\ell$ for simplicity. The emergence of Kasner eons is evident, exhibiting the predicted values, although the cubic eon in $d=5$ is absent. This absence arises because, in $d \geq 5$, cubic quasi-topological gravities behave like cubic Lovelock gravity, which is topological in $d=5$.}
     \label{fig:1}
\end{figure}

In Figure \ref{fig:1}, we present the effective Kasner exponents \eqref{eq:effkasnervac} for specific coupling choices $\lambda_n$ in \eqref{eq:laggen} across $d=4$ and $d=5$ spatial dimensions. Each graph clearly displays a sequence of Kasner eons, represented by the flat regions, followed by rapid transitions to the next eon. Additionally, we find that each Kasner eon exhibits the predicted values \eqref{eq:kasnergenbhvac} for $p_t$ and $p_x$.

\subsection{Kasner eons with matter}
\label{subsec:mat}

We now aim to investigate how matter fields affect Kasner eons. As an illustrative example to understand the interplay between higher-curvature terms and matter, we will consider a minimally coupled scalar field on top of our gravity theory \eqref{eq:laggen}:
\begin{equation}
    \Lag=\frac{\lambda_0}{\ell ^2}+ \lambda_1 R+\sum_{n=2}^\infty \lambda_n \ell^{2(n-1)} \mathcal{Z}_{(n)}-\frac{1}{2} (\partial \phi)^2-V(\phi)\,.
    \label{eq:laggens}
\end{equation}
At this point, it is natural to wonder about including higher-derivative terms in the scalar field $\phi$, as well as non-minimal couplings between gravity and the scalar field. However, a technical challenge arises: the resulting equations of motion, even within the highly symmetric configurations described by \eqref{eq:bbans}, would generally introduce higher derivatives in the fields, complicating the numerical integration. Therefore, in this study, we will limit our analysis to the inclusion of the standard two-derivative kinetic term, which will already introduce significant new physical features, as we will demonstrate below.

After incorporating the minimally coupled scalar field, the equations of motion for the resulting theory \eqref{eq:laggens} are as follows:
\begin{align}
     P_{acde} R_b{}^{cde}-\frac{1}{2} g_{ab} \mathcal{L}+2 \nabla^{c} \nabla^{d} P_{acbd}&=\frac{1}{2} \partial_a \phi\, \partial_b \phi\,, \qquad P^{abcd}\equiv\frac{\partial \mathcal{L}}{\partial R_{abcd}}\,, \\ \Box\,  \phi &= V'(\phi)\,.
\end{align}
where $\mathcal{L}$ now refers to \eqref{eq:laggens}. As before, we assume a plane-symmetric ansatz for the system, so the metric and the scalar field are written as:
\begin{equation}
\label{eq:bbansscal}
    \mathrm{d}s^2=\frac{1}{r^2} \left[ - e^{-\chi(r)}f(r) \diff t^2+\frac{1}{f(r)}\diff r^2+ \mathrm{d} x_{d-1}^2 \right]\,, \qquad \phi=\phi(r)\,.
\end{equation}
Under this assumption, the condition $\nabla^d P_{acbd} \vert_{f,\chi}=0$ continues to hold. This simplification allows the complete set of equations of motion to be reduced to a system of three ordinary differential equations, each containing no more than two derivatives for $f(r)$, $\chi(r)$ and $\phi(r)$. These equations take the following form:
\begin{align}
\label{eq:eomscalar}
r^2 f \phi ''&=\frac{1}{2} r \, \phi' \left(f \left(r \chi '+2(d-1)\right)-2 r f'\right)+V'(\phi(r))\,.\\
\label{eq:chiprimephi}
\chi'&=\frac{r (\phi')^2}{\sum_{n=1}^\infty \tilde{\lambda}_n \ell^{2(n-1)} f^{n-1}}\,, \\
f'&=\frac{-2\lambda_0+2\ell^2 V(\phi(r))+2d\sum_{n=1}^\infty \frac{\ell^{2n} \tilde{\lambda}_n}{n} f^n+  \ell^2 f r^2 (\phi')^2 } {2r \sum_{n=1}^\infty \ell^{2n} \tilde{\lambda}_n f^{n-1}}\,,
\label{eq:eomfprimephi}
\end{align}
where $\tilde{\lambda}_n$ are dimensionless constants given by:
\begin{equation}\label{rescaledlambdas}
    \tilde{\lambda}_n=(-1)^{n+1} \, n (d+1-2n) (d(d+1))^{n-1} \lambda_n\,.
\end{equation}
Similar to the vacuum case, our focus is on determining the Kasner exponents associated with each Kasner eon that emerges as we approach the black hole singularity. To achieve this, we consider solutions in regions where they can be locally approximated by pure Kasner solutions. This approximation holds when $f(r)$, $\chi(r)$ and $\phi(r)$ exhibit the following profiles:
\begin{equation}
    \!f(r)=-f_n r^{\gamma_n}\,,\qquad \chi(r)= \chi_n \log r+\chi_0\,, \qquad \phi(r)=\begin{cases} \phi_1 \log r \quad \mathrm{(GR)}\,, \\ \phi_n r^{\nu_n}  \quad  \mathrm{(higher}\text{-}\mathrm{curvature)}\,,
    \end{cases}
    \label{eq:kasneransbbscalarr}
\end{equation}
where we explicitly distinguish between the logarithmic profile of the scalar field in Einstein gravity \cite{Damour:2002et} and its power-law behavior in the presence of higher-curvature corrections. In analogy with the vacuum case, we define a new coordinate $f_n r^{\gamma_n+2}\mathrm{d}\tau^2=\mathrm{d}r^2$ so that, with the profiles for $f(r)$ and $\chi(r)$ given in \eqref{eq:kasneransbbscalarr}, \eqref{eq:bbansscal} transforms into a pure Kasner metric:
\begin{equation}
    \mathrm{d}s^2=-\mathrm{d}\tau^2+\tau^{2 p_t} \mathrm{d} t^2+ \tau^{2 p_x} \mathrm{d} x^2_{d-1}\,,\qquad p_x=\frac{2}{\gamma_n}\,, \qquad p_t=p_x+\frac{\chi_n}{\gamma_n}-1\,.
    \label{eq:kasnergenbh}
\end{equation}
Now, observe that any given curvature invariant $\mathcal{R}_{(n)}$ of order $n$, when evaluated on a Kasner metric \eqref{eq:kasner}, diverges in the limit $\tau \rightarrow 0$ as:
\begin{equation}
  \mathcal{R}_{(n)} \sim \frac{1}{\tau^{2n}}\,.  
\end{equation}
If we restrict our attention to quadratic potentials of the form $V(\phi)=\tfrac{1}{2}m^2\phi^2$, then the kinetic term will always diverge faster than $V(\phi)$. Therefore, for the scalar field to influence the Kasner eon, its kinetic terms must match the order of $\mathcal{R}_{(n)}$. This implies that the scalar field must diverge as $1/\tau^{n-1}\propto r^{(n-1)\gamma_n/2}$, so that:\footnote{This analysis is applicable only to higher-order eons with $n\geq2$. In the case of Einstein gravity, where $n=1$, the scalar field exhibits the well-known logarithmic profile (\ref{eq:kasneransbbscalarr}).}
\begin{equation}
    \nu_n=\frac{(n-1)\gamma_n}{2}\,.
    \label{eq:expscalar}
\end{equation}
The Kasner exponent associated with the scalar, $p_\phi$, can be naturally defined such that $\phi(\tau)\propto \tau^{p_\phi}$ as $\tau\to0$. 
This definition implies
\begin{equation}
    p_\phi=-\frac{2}{\gamma_n} \nu_n=-n+1\,.
    \label{eq:kasnerexpscalarfield}
\end{equation}
Next, let us investigate the values of the remaining Kasner exponents $p_t$ and $p_x$ in each eon. To do this, consider a regime where the dominant contributions to the equations of motion are the terms of order $O(\ell^{2(n-1)})$, for $n>1$. This entails imposing \eqref{eq:kasneransbbscalarr} on the equations of motion and disregarding any subdominant terms. Furthermore, if the potential is quadratic in $\phi$, it can be ignored, as its contribution is subleading compared to the kinetic term. Consequently, the scalar equation of motion \eqref{eq:eomscalar} implies that:
\begin{equation}
    (n+1)\gamma_n-\chi_n=2d\,.
\end{equation}
On the other hand, if we solve for $\phi'$ in \eqref{eq:eomfprimephi}, substitute it in \eqref{eq:chiprimephi}, impose \eqref{eq:kasneransbbscalarr} and disregard terms arising from curvature invariants of different order than $n$, we find:
\begin{equation}
    -n \chi_n+2n\gamma_n=2d\,.
\end{equation}
By combining the previous two equations, we arrive at:
\begin{equation}
\gamma_n=\chi_n=\frac{2d}{n}\,.
\end{equation}
Using \eqref{eq:kasnergenbh}, this implies that the exponents $p_t$ and $p_x$ for the $n$-th Kasner eon are:
\begin{equation}
    p_t=p_x=\frac{n}{d}\,.
    \label{eq:kasnerexpscalarmet}
\end{equation}
These Kasner exponents, together with \eqref{eq:kasnerexpscalarfield}, modify the Kasner relations as follows:\footnote{These relations assume $\lambda_1=1$ and a canonically normalized scalar field. Note that the seminal reference \cite{Damour:2002et} uses a different normalization for the scalar.}
\begin{equation}
    \sum_{i=1}^d p_i=n\,, \qquad \sum_{i=1}^d p_i^2+\frac{1}{2}p_\phi^2=\frac{n^2}{d}+\frac{(n-1)^2}{2}\,.
    \label{eq:kasnerrelnewscalar}
\end{equation}
Let us now comment on these results. First, note that naively setting $n=1$ in \eqref{eq:kasnerrelnewscalar} does not yield the correct relations for Einstein gravity with a minimally coupled scalar field, where $\sum_{i=1}^d p_i=1$ and $\sum_{i=1}^d p_i^2+\frac{1}{2}p_\phi^2=1$. This difference is expected, as the scalar field in Einstein gravity exhibits a logarithmic divergence $\phi(\tau)\sim p_\phi \log \tau$, in contrast to the power-law behavior in higher-order cases, which results in a distinct definition for $p_\phi$.

Second, while the Kasner exponents $p_x$ and $p_t$ were distinct in the absence of the scalar field \eqref{eq:kasnerexpeonvac}, introducing the scalar field dramatically changes the situation. Now, the Kasner exponents for any Kasner eon ---except for the unique case of Einstein gravity with matter--- become equal. Interestingly enough, this means that the presence of the scalar field enhances the isotropy of the Kasner solution, effectively rendering the temporal direction (i.e., that associated with the coordinate $t$) indistinguishable from any of the spatial directions. One may interpret that matter produces an additional \emph{mixing} between the Kasner directions that render them identical. The price to pay by that \emph{scrambling} is that now Kasner exponents are halved compared to the $p_x$ exponent in purely gravitational higher-curvature theories. Additionally, the shift $\delta p$ in Kasner exponents between successive eons is also halved in the presence of the scalar field:
\begin{equation}
    \delta p=\frac{1}{d}\,.
    \end{equation}
    
There is also an important difference between GR and higher-curvature theories when incorporating a minimally coupled scalar field. In the presence of higher-curvature corrections, the Kasner exponents associated with both the spacetime metric and the scalar field are fully determined by \eqref{eq:kasnerexpscalarfield} and \eqref{eq:kasnerexpscalarmet}. In contrast, the Kasner exponents are less constrained in Einstein gravity with a minimally coupled scalar field. Specifically, if we examine the equations of motion \eqref{eq:eomscalar}-\eqref{eq:eomfprimephi} and focus on the Einstein eon, imposing \eqref{eq:kasneransbbscalarr} led us to:
\begin{equation}
    \chi_1=\frac{\phi_1^2}{(d-1)}\,,\qquad  \gamma_1=d+\frac{\phi_1^2}{2(d-1)}\,.
\end{equation}
This implies that the set of Kasner exponents in the GR eon is given by:
\begin{equation}
    p_x=\frac{4(d-1)}{2d(d-1)+\phi_1^2}\,,\qquad p_t=\frac{\phi_1^2-2(d-1)(d-2)}{2d(d-1)+\phi_1^2}\,, \qquad p_\phi=-p_x \phi_1\,.
    \label{eq:kasnerexpgrbh}
\end{equation}
Note that the relationships $p_t+(d-1)p_x=1$ and $p_t^2+(d-1)p_x^2 +\frac{1}{2}p_\phi^2=1$ hold true, as expected. Consequently, we find that the Kasner exponents in the GR eon are not fully determined; there is one free parameter that can be identified as either $\phi_1$ or $p_\phi$. In contrast, the set of Kasner exponents for higher-curvature eons is fully determined, which can possibly be attributed to the lack of higher-derivative terms in the scalar field. In this case, we needed to select the appropriate power for the scalar field to ensure it matches the order of the higher-curvature terms ---see \eqref{eq:expscalar}--- thereby eliminating the free parameter present in GR.
    
This concludes the analytical characterization of the various Kasner eons associated with static, plane-symmetric solutions to \eqref{eq:laggens}. We will now turn our attention to exploring these features through explicit numerical examples.

To begin with, let us provide some details about the numerical treatment of \eqref{eq:eomscalar}-\eqref{eq:eomfprimephi}. As in the vacuum case, we pick a negative $\lambda_0$ to guarantee AdS asymptotics and planar black hole solutions, preparing for the holographic studies in section \ref{sec:hol}. However, we emphasize that the cosmological constant does not play any role in the approach towards the singularity. Now, by assuming regularity at the horizon, $r=r_h$, we may locally expand $\phi$, $f$ and $\chi$ as:
\begin{align}
\phi(r) &= \phi_h + \phi'_h (r-r_h) + O[(r-r_h)^2]\,,\\
f(r) &= f'_h (r-r_h) + O[(r-r_h)^2]\,,\\
\chi(r) &= \chi_h + \chi'_h (r-r_h) + O[(r-r_h)^2]\,,
\end{align}
where the subindex $h$ indicates evaluation at the horizon. Further, by evaluating the field equations at the horizon, we find the following relations between the variables:
\begin{equation}
    r_h^2 \phi'_h f'_h=V'(\phi_h)\,, \qquad (d-1) \chi'_h=r_h (\phi_h')^2\,, \qquad 2 \ell^2 r_h (d-1) f'_h=-2\lambda_0 +2 \ell^2 V(\phi_h)\,,
    \label{eq:consiv}
\end{equation}
 Once the potential $V(\phi)$ is specified, one can solve for $\chi_h'$, $\phi_h$ and $\phi_h'$ in terms of $r_h$ and $f'_h$. The latter is related to the black hole temperature $T$ by $4 \pi T= -f'_h e^{-\chi_h/2}$. Notably, the constraints given by \eqref{eq:consiv} do not depend on the higher-curvature terms.

We then proceed with the numerical integration by initializing the algorithm at a point $r=r_h\pm\delta$ sufficiently close to the horizon, with $\delta \ll 1$. Additionally, we set $\chi_h=0$, which can always be accomplished via a time reparametrization. Consequently, by fixing $f'_h$ and $r_h$ and specifying a value for $\delta$, we are able to solve for the region between the horizon and the singularity, and between the horizon and the AdS boundary, respectively.

To efficiently visualize the different Kasner eons as we approach the singularity, we now introduce the effective Kasner exponents:
\begin{equation}
    p_x^{\rm eff}=\frac{2 f(r)}{r f'(r)}\,, \quad  p_t^{\rm eff}=\frac{2 f(r)}{r f'(r)}+\frac{f(r) \chi'(r)}{f'(r)}-1\,, \quad  p_\phi^{\rm eff}(r)=-\frac{2 \chi'(r) \phi'(r)}{ \phi(r)} \left ( \frac{f'(r)}{f(r)} \right)^2 \,.\label{effpswithmatter}
\end{equation}
Indeed, when $f(r), \chi(r)$ and $\phi(r)$ match a proper Kasner eon near the singularity \eqref{eq:kasneransbbscalarr}, the effective exponents $p_x^{\rm eff}$, $p_t^{\rm eff}$ and $p_\phi^{\rm eff}$  are given by  \eqref{eq:kasnerexpscalarmet} and \eqref{eq:kasnerexpscalarfield}. Furthermore, we explicitly include the factor $\chi'(r)$ in $p_\phi^{\rm eff}(r)$ since it serves as a clear indicator of the scalar field's presence ---whenever $\chi$ is constant, leading to $\chi'=0$, the scalar field profile becomes trivial, cf. \eqref{eq:chiprimephi}. These expressions extend the effective Kasner exponents given in \cite{Bueno:2024fzg} to cases where the scalar field is present and $\chi(r)$ is non-constant. 

\begin{figure}[t!]
\includegraphics[scale=0.33]{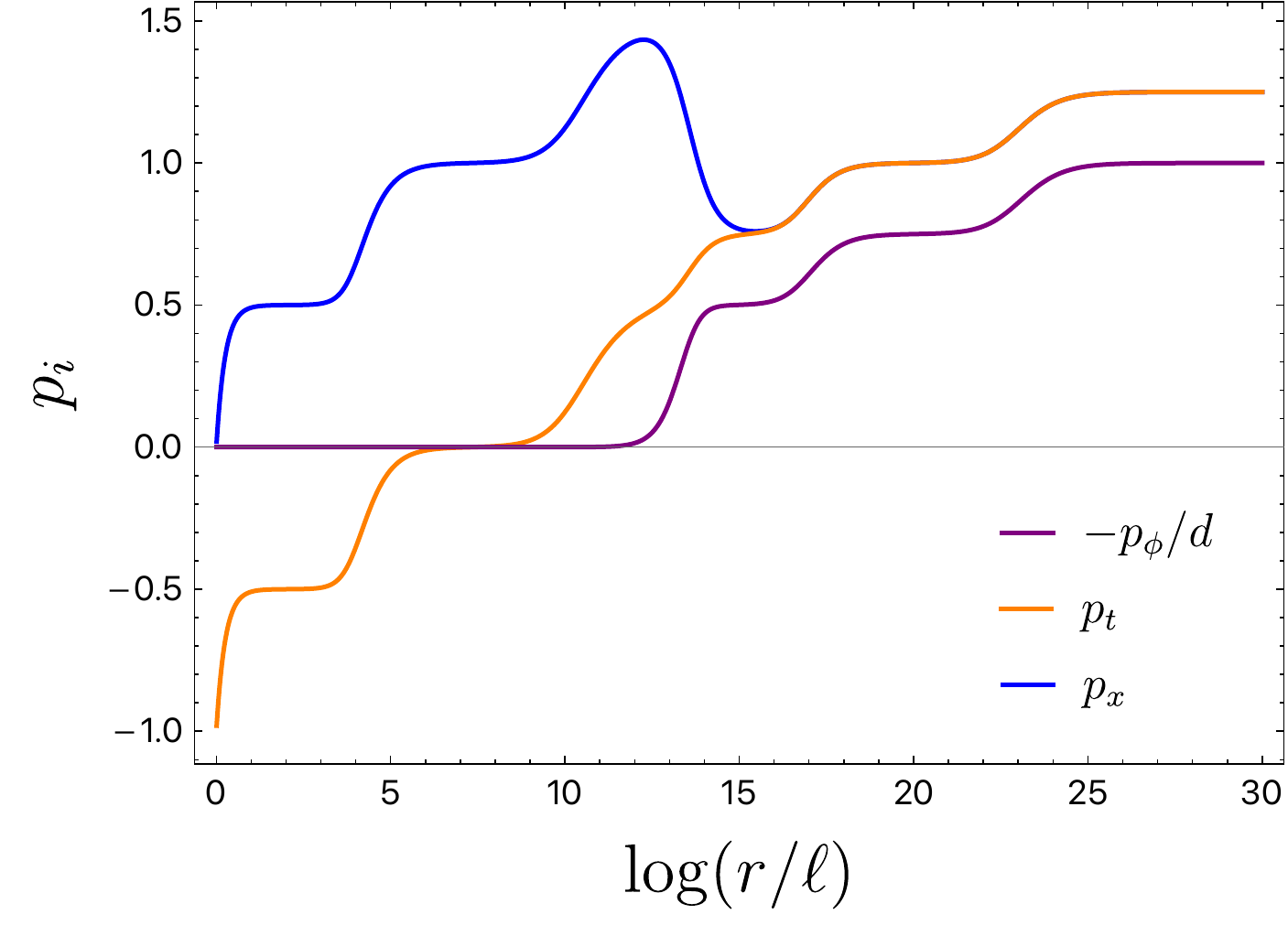}
     \includegraphics[scale=0.33]{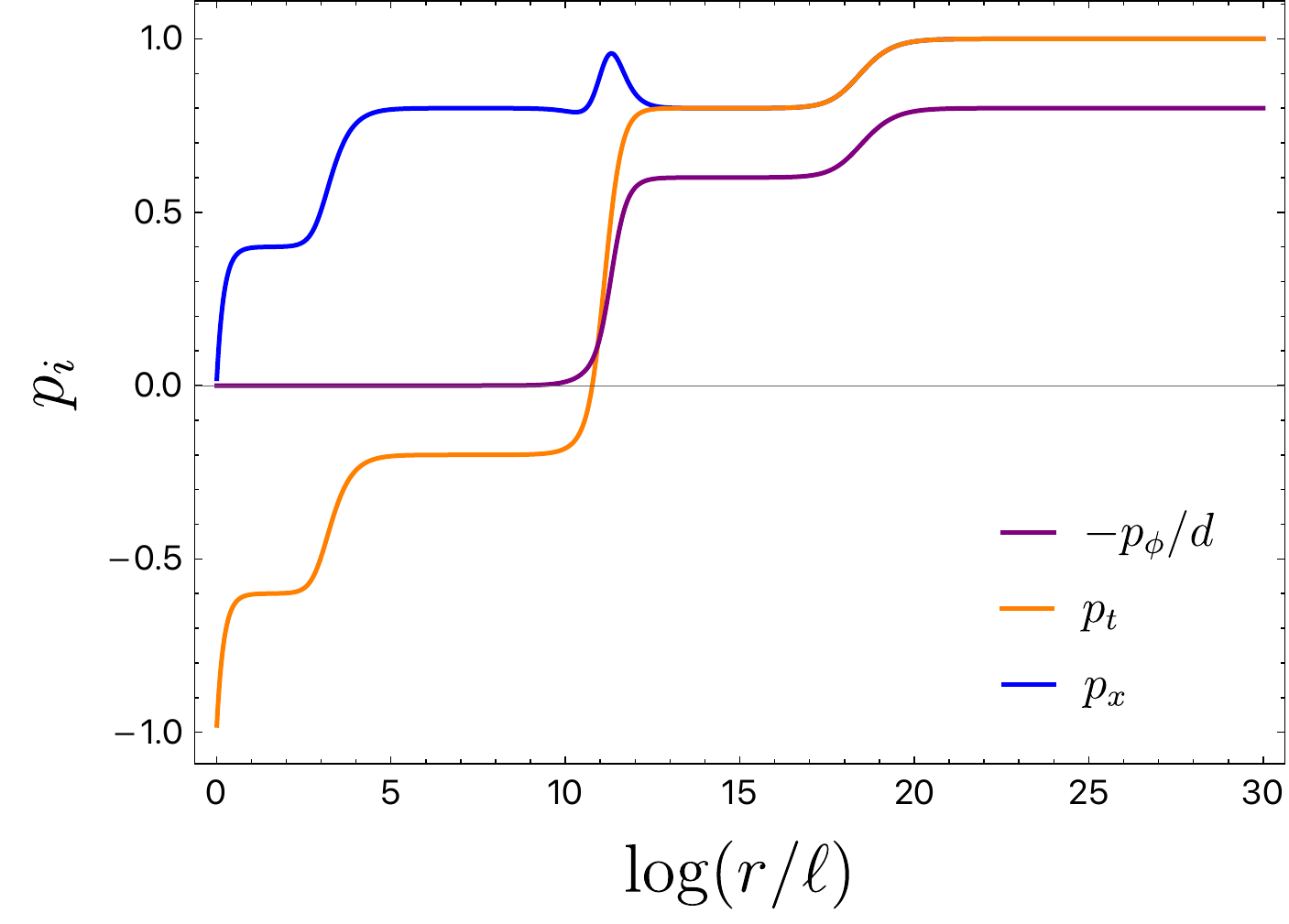}
     \caption{Effective Kasner exponents for higher-curvature gravities \eqref{eq:laggen} in $d=4$ (left) and $d=5$ (right) with the same specific dimensionless couplings $\lambda_n$ used in Figure \ref{fig:1} ($\lambda_0=d(d-1)$, $\lambda_1=1$, $\lambda_2=10^{-8}$, $\lambda_3=-10^{-22}$, $\lambda_4=-10^{-42}$, $\lambda_5=-10^{-67}$ and $\lambda_{n>5}=0$) and $m^2=-10^{-4}/\ell^2$. As reviewed in Appendix \ref{app:holdic}, a negative $m^2$ is consistent with the Breitenlohner-Freedman stability bound \cite{Breitenlohner:1982bm,Breitenlohner:1982jf}. We have also set $r_h=\ell$ for simplicity. The expected Kasner eons emerge with the predicted values, and the effective Kasner exponent $p_\phi^{\rm eff}$ is rescaled by a factor $-1/d$ for clarity in the plots.}
     \label{fig:2}
\end{figure}

Figure \ref{fig:2} shows the effective Kasner exponents for a specific set of couplings and a massive free scalar field with $V(\phi)=\frac{1}{2} m^2 \phi^2$. Notably, the initial Kasner eon is driven by the Einstein gravity eon with a massive scalar field. While this feature cannot be directly observed from Figure \ref{fig:2} ---as the GR eon produces a logarithmic profile for the scalar field, which $p_\phi^{\rm eff}$does not capture--- we have confirmed that the numerical solution within the interval defining the GR eon satisfies the conditions \eqref{eq:kasnerexpgrbh}, characteristic of the Einstein gravity eon with a scalar field. Subsequently, the next eon is dominated by a purely gravitational Gauss-Bonnet term, with no significant influence from the scalar field. This behavior suggests that the scalar field requires some time to transition from a logarithmic approach to the singularity to a power-law behavior. Depending on the value of the scalar field at the black hole horizon, this transient period may persist for an extended period, during which the system may only experience transitions between purely gravitational Kasner eons.

Eventually, the scalar field reaches its characteristic power-law divergence, completing the transition to Kasner eons influenced by the minimally coupled scalar field. In the plots shown in Figure \ref{fig:2}, this transition occurs as the purely gravitational quadratic Kasner eon gives way to the next eon. From that point onwards, the scalar field becomes significant enough to impact the Kasner dynamics. The scalar field and higher-curvature terms then become closely intertwined, evolving in accordance with \eqref{eq:kasnerexpscalarfield} and \eqref{eq:kasnerexpscalarmet}.

\section{Imprints of Kasner eons on holographic observables}
\label{sec:hol}

In this section, we introduce several holographic probes to identify the presence of Kasner eons. Our focus will be on holographic CFTs dual to higher-curvature gravitational theories that include a minimally coupled massive scalar:
\begin{equation}
\label{eq:lagholo}
I = \frac{1}{16 \pi G_{d+1}} \int d^{d+1} x \sqrt{-g} \left( R + d(d-1) + \sum_{n=2}^\infty \lambda_n \mathcal{Z}_{(n)}-\frac{1}{2} (\partial \phi)^2 - \frac{1}{2} m^2 \phi^2 \right).
\end{equation}
These theories were examined from a purely gravitational perspective in section \ref{subsec:mat}. In fact, the action \eqref{eq:lagholo} follows from \eqref{eq:laggens} with $\lambda_0=d(d-1)$, $\ell=1$ and $V(\phi) = \frac{1}{2} m^2 \phi^2$. 

We begin by considering the following plane-symmetric ansatz,
\begin{equation}
\mathrm{d}s^2 = \frac{1}{r^2} \left( -f(r) e^{-\chi(r)} \mathrm{d}t^2 + \frac{\mathrm{d}r^2}{f(r)} + \mathrm{d} \vec{x}_{d-1}^2 \right)\,, \qquad \phi=\phi(r)\,.
\label{metAnsatz}
\end{equation}
where $\mathrm{d} \vec{x}_{d-1}^2$ denotes the $(d-1)$-dimensional Euclidean metric. In this coordinate system, the conformal boundary is at $r = 0$, while the singularity lies at $r \to \infty$. The bulk equations of motion, along with their numerical treatment, were thoroughly discussed in section \ref{subsec:mat}. It was noted that solving the equations numerically from the horizon towards either the asymptotic boundary or the singularity required setting initial conditions for the scalar field, its derivative, and the metric function $\chi(r)$, cf. \eqref{eq:consiv}. In the AdS/CFT correspondence, the scalar field is dual to a scalar operator $\mathcal{O}$ with a conformal dimension $\Delta$, satisfying:\footnote{See Appendix \ref{app:holdic} for more details on the holographic relations between $\phi$ and $\mathcal{O}$.}
\begin{equation}\label{massdelta}
    m^2=\Delta(\Delta-d)\,.
\end{equation}
In terms of $\Delta$, the initial conditions \eqref{eq:consiv} take the form:
\begin{align}
\phi_h = \mp \frac{\sqrt{2}\sqrt{d-1}\sqrt{-d-f'_h r_h}}{\sqrt{\Delta(d-\Delta)}}\,,&\qquad 
\phi'_h = \pm\frac{\sqrt{2}\sqrt{d-1}\sqrt{-d-f'_h r_h}\sqrt{\Delta(d-\Delta)}}{f'_h r_h^2}\,,\\
\chi'_h &= -\frac{2(d+f'_h r_h)\left[\Delta(d-\Delta)\right]}{f'^2_h r_h^3}\,.
\end{align}
It is important to emphasize that hairy black holes with non-trivial scalar profiles correspond to RG flows in the dual field theory, where the putative CFT is deformed by the operator $\mathcal{O}$. In this context, it is essential to ensure that $\Delta<d$, so that the dual operator corresponds to a relevant deformation (thereby, affecting the IR of the theory). According to \eqref{massdelta}, this implies we must select a negative $m^2$, however, in AdS, there exists an allowable range that complies with the Breitenlohner-Freedman stability bound \cite{Breitenlohner:1982bm,Breitenlohner:1982jf}.\footnote{For positive $m^2$ the dual operator corresponds to an irrelevant information. In this case, the effects of the deformation are mostly noticed near the boundary, which ceases to be AdS. Conversely, these effects become negligible in the deep interior, indicating that they will not influence the near-singularity region.}

With these preliminaries in place, we now turn to examining several holographic quantities that can reveal the presence of Kasner eons from the field theory point of view.

\subsection{Thermal $a$-function}

In the context of quantum field theories (QFTs), the degrees of freedom decrease due to coarse-graining along the renormalization group flow. The existence of a monotonic function that quantifies the degrees of freedom is well established in two dimensions \cite{Zamolodchikov:1986gt}. This issue has also been explored in higher dimensions \cite{Komargodski:2011vj, Komargodski:2011xv, Casini:2004bw, Casini:2012ei, Casini:2017vbe}, where the monotonic function is known as the $a$-function. Via the standard UV/IR relations in holography, it is well known that the radial direction maps to an energy scale in the field theory \cite{Susskind:1998dq,Peet:1998wn,Hatta:2010dz,Agon:2014rda}. This indicates that the holographic dual of the $a$-function should be a bulk quantity that decreases from the boundary (UV) to the horizon (IR).  Various proposals for a holographic $a$-function have been proposed and thoroughly investigated in the literature \cite{Freedman:1999gp, Myers:2010xs, Myers:2010tj,GonzalezLezcano:2022mcd,Deddo:2023pid}. More recently, a finite temperature generalization of this function, dubbed the thermal $a$-function, was introduced in \cite{Caceres:2022smh} to probe and characterize black hole interiors within holographic RG flows. This quantity has been further explored in various contexts in \cite{Caceres:2023mqz, Caceres:2022hei, Caceres:2023zft, Arean:2024pzo,Carballo:2024hem}.

One effective approach to ensure the monotonicity of the $a$-function is to require that the matter fields satisfy the Null Energy Condition (NEC),
\begin{equation}\label{eq:nec}
k^\mu k^\nu T_{\mu\nu} \geq 0\,,
\end{equation}
for any null vector \(k^\alpha\) and the corresponding bulk stress-energy tensor \(T_{\mu\nu}\).
Alternatively, in higher-curvature gravities, we can impose the Null Curvature Condition (NCC), which stipulates that the Ricci curvature of a Lorentzian manifold $R_{\mu\nu}$ must not be negative along any arbitrary null vector,\footnote{In Einstein gravity, these two conditions are equivalent. However, in the context of higher-curvature gravities, the distinction roughly depends on how we interpret the additional terms in the field equations ---whether as matter contributions or as modifications to the geometric sector.}
\begin{equation}\label{eq:ncc}
    k^{\mu}k^{\nu}R_{\mu\nu}\geq 0\,.
\end{equation}
Thus, \eqref{eq:nec} and \eqref{eq:ncc} offer two distinct approaches to defining a monotonic function along the holographic flow. Below, we will propose two monotonic functions corresponding to these conditions and explicitly use them to prove their monotonicity.

\begin{figure}[t!]\centering
    \begin{subfigure}{0.45\textwidth}
        \includegraphics[width=\hsize,trim={0 0 -1.1cm 0},clip]{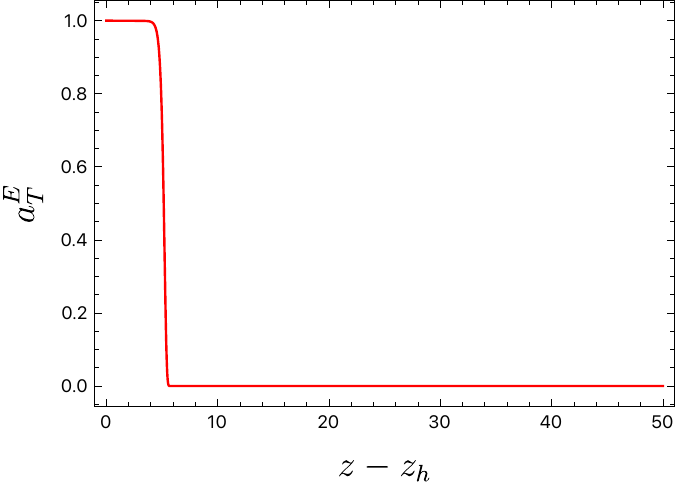}
        \captionsetup{justification=centering}
       \caption{}
        \label{fig:ae}
    \end{subfigure}
\hfil
    \begin{subfigure}{0.44\textwidth}
    \includegraphics[width=\hsize,trim={0 0 -1.1cm 0},clip]{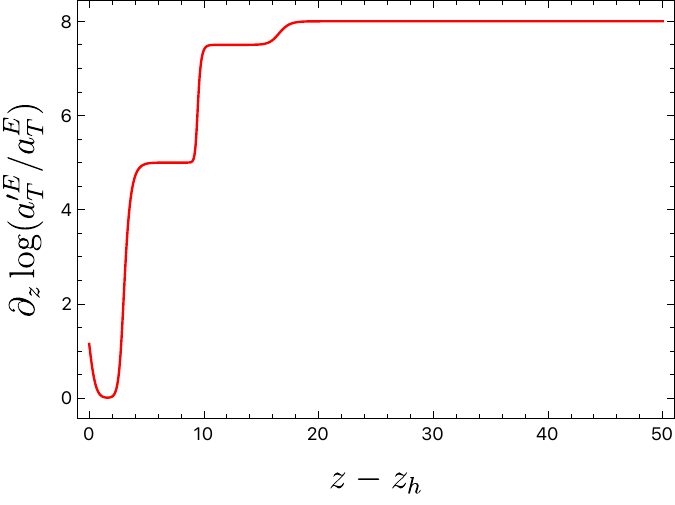}
    \captionsetup{justification=centering}
   \caption{}
    \label{fig:eons_ae}
\end{subfigure}

    \begin{subfigure}{0.45\textwidth}
        \includegraphics[width=\hsize,trim={0 0 -1.1cm 0},clip]{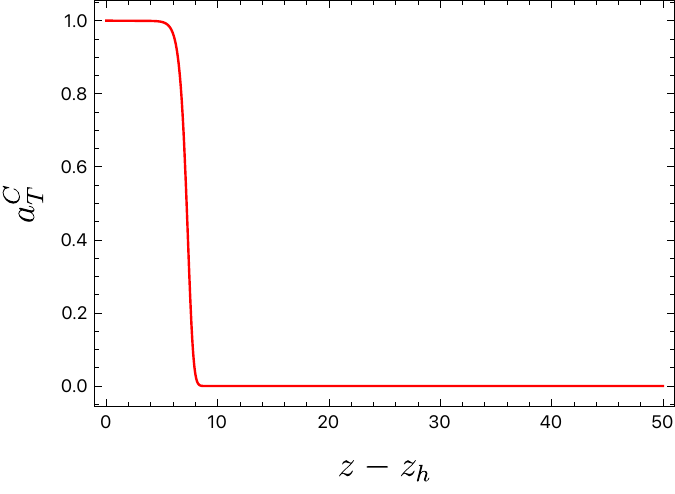}
        \captionsetup{justification=centering}
        \caption{}
        \label{fig:ac}
    \end{subfigure}
\hfil
    \begin{subfigure}{0.44\textwidth}
    \includegraphics[width=\hsize,trim={0 0 -1.1cm 0},clip]{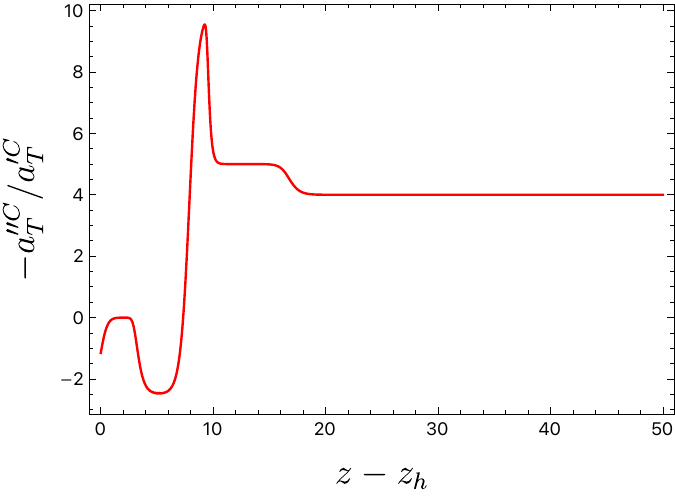}
    \captionsetup{justification=centering}
    \caption{}
    \label{fig:eons_ac}
\end{subfigure}
\vspace{-3mm}
    \caption{Thermal $a$-functions in units of $\pi^{d/2}/\ell_P^{d-1}\Gamma \left(\frac{d}{2}\right)$ (left) and particular combinations of their first and second derivatives constructed to identify the eons (right) in $d=5$ with $z=\log (r)$, $\lambda_2=10^{-8}$, $\lambda_3=-10^{-22}$, $\lambda_4=-10^{-42}$, $\lambda_5=-10^{-67}$, $\lambda_{n>5}=0$, and $m^2=-\frac{d^2}{4}+\frac{1}{2}$.}
    \label{fig:ae_ac}
\end{figure}

Working with the ansatz \eqref{metAnsatz}\footnote{The thermal $a$-function was initially defined using the domain wall ansatz and analytically continued into the black hole interior \cite{Caceres:2022smh}. However, this explicit analytic continuation is unnecessary for the coordinates used in \eqref{eq:bbansscal}. The continuation is naturally implemented by the radial coordinate $r$, which transitions to a time-like direction in the interior of the black hole.}
and imposing the NEC \eqref{eq:nec} yields a monotonic function, which we denote as $a^{\text{E}}_T(r)$,
\begin{equation}
a^{\text{E}}_T(r)= \frac{\pi^{d/2}}{\Gamma \left(\frac{d}{2}\right)} \left(\frac{1}{\ell_P}\right)^{d-1}\exp\bigg(-\frac{(d-1)}{2}\chi-\frac{1}{2}\sum_{n=2}^\infty \tilde{\lambda}_n\int  f^{(n-1)}\chi'dr\bigg)\,,\label{afunction1}
\end{equation}
where $\tilde{\lambda}_n$ are the rescaled couplings introduced in \eqref{rescaledlambdas}. Alternatively, by imposing the NCC \eqref{eq:ncc}  we obtain a function that we denote as $a^{\text{C}}_T(r)$,
\begin{equation}
a^{\text{C}}_{T}(r)=\frac{\pi^{d/2}}{\Gamma \left(\frac{d}{2}\right)} \left(\frac{1}{\ell_P}\right)^{d-1}\exp\bigg(-\frac{(d-1)}{2}\chi\bigg)\,.
\label{afunction2}
\end{equation}
Monotonicity of $a^{\text{E}}_T(r)$ and $a^{\text{C}}_T(t)$ can be demonstrated by analyzing their derivative along the entire flow.\footnote{The standard RG flow describes the region outside the black hole, while its `trans-IR' extension corresponds to the interior of the black hole.} In particular, we find that
\begin{eqnarray}
\frac{da^{\text{E}}_T}{dr} =-\frac{a^{\text{E}}_T}{2r f(r)^2}\bigg[f(r)\big(T^{r}_{\;r} - T^{t}_{\; t}\big)\bigg]\,,
\end{eqnarray}
and
\begin{eqnarray}
\frac{da^{\text{C}}_T}{dr} =-\frac{a^{\text{C}}_T}{2r f(r)^2}\bigg[f(r)\big(R^{r}_{\;r} - R^{t}_{\; t}\big)\bigg]\,,
\end{eqnarray}
respectively, where 
\begin{eqnarray}
T^{r}_{\;r} - T^{t}_{\; t} &=&r f(r)\bigg(\chi'{(d-1)+\sum_{n=2}^\infty \tilde{\lambda}_n f^{(n-1)}\chi'}\bigg)\,, \\
R^{r}_{\;r} - R^{t}_{\; t} &=&{(d-1)}r f(r)\chi'\,.
\end{eqnarray}
On the other hand, using the null vector \(\vec{k} = e^{-\chi(r)/2}\partial_t + f(r)\partial_r\), we can rewrite the NEC and NCC as follows:
\begin{equation}
f(r)\big(T^{r}_{\;r} - T^{t}_{\; t}\big)  \geq 0\,,\qquad f(r)\big(R^{r}_{\;r} - R^{t}_{\; t}\big) \geq 0\,,
\end{equation}
implying \(da^{\text{E}}_T / dr \leq 0\) and $da^{\text{C}}_T / dr \leq 0$ everywhere along the flow. Thus, both thermal $a$-functions monotonically decrease from the AdS boundary to the singularity. We illustrate this behavior in Figures \ref{fig:ae} and \ref{fig:ac}.\footnote{Note that in the absence of the scalar deformation, $\chi$ is trivial and both $a$-functions remain constant, as they should. Thus, by definition, they are unable to probe vacuum eons.}

We observe several noteworthy properties of these functions. First, when higher-curvature corrections are absent, $a_T^{\text{E}}$ and $a_{T}^{\text{C}}$ coincide. This is expected, as the NEC and NCC are equivalent in Einstein's gravity, which is no longer true when higher-curvature corrections are introduced. Second, both functions effectively capture \emph{all} matter eons and their transitions through particular combinations of their first and second derivatives. For example, note that
\begin{align}\label{aefunceons}
\!\!\!\partial_z\log\bigg[{\frac{a'^{E}_T(z)}{a^{E}_T(z)}}\bigg]=\frac{2p'^{\text{eff}}_\phi(z)}{p^{\text{eff}}_\phi(z)}+\frac{2(p^{\text{eff}}_\phi(z)-p^{\text{eff}}_x(z)p'^{\text{eff}}_t(z)-(1+p^{\text{eff}}_t(z)-2p^{\text{eff}}_x(z))p'^{\text{eff}}_x(z)}{(1+p^{\text{eff}}_t(z)-p^{\text{eff}}_x(z))p^{\text{eff}}_x(z)}\,,&\\
\!\!\!-\frac{a''^{C}_T(z)}{a^{C}_T(z)}=\frac{(d-1) \left(p^{\text{eff}}_t(z)-p^{\text{eff}}_x(z)+1\right){}^2-p^{\text{eff}}_x(z)
   p_t'^{\text{eff}}(z)+\left(p^{\text{eff}}_t(z)+1\right) p_x'^{\text{eff}}(z)}{p_x^{\text{eff}}(z)
   \left(p_t^{\text{eff}}(z)-p_x^{\text{eff}}(z)+1\right)}\,,&\label{acfunceons}
\end{align}
where $p^{\text{eff}}_t(z)$, $p^{\text{eff}}_x(z)$ and $p^{\text{eff}}_\phi(z)$ are given in \eqref{effpswithmatter} and we have defined $z=\log r$ for convenience.
During each Kasner eon, $p^{\text{eff}}_t(z)$, $p^{\text{eff}}_x(z)$ and $p^{\text{eff}}_\phi(z)$ stabilize at a plateau, indicating that (\ref{aefunceons}) and (\ref{acfunceons}) must reach constant values as well:
\begin{align}
    \partial_z\log\bigg[{\frac{a'^{E}_T(z)}{a^{E}_T(z)}}\bigg]&\to\frac{2 p_{\phi }}{p_x \left(p_t-p_x+1\right)}\,,\\
    -\frac{a''^{C}_T(z)}{a^{C}_T(z)}&\to\frac{(d-1) \left(p_t-p_x+1\right)}{p_x}\,.
\end{align}
This behavior is depicted in Figures \ref{fig:eons_ae} and \ref{fig:eons_ac}, demonstrating that these monotonic functions serve as a robust diagnostic tool for identifying stringy effects near black hole singularities. Finally, we note that, among the two functions, only $a^{\text{E}}_T$ defined in \eqref{afunction1}, approaches the zero temperature $a$-function proposed in \cite{Myers:2010jv} for quasi-topological gravities with $\lambda_{n\geq 4}=0$.
To see this, we switch to the conventional domain wall coordinates,
\begin{equation}
    ds^2=e^{2A(\rho)}(-h(\rho)^2dt^2+d\vec{x}_{d-1}^2)+d\rho^2\,,
\end{equation}
where
\begin{equation}
e^{2A(\rho)} = \frac{1}{r^2}\,, \qquad h(\rho)^2 = f(r)e^{-\chi(r)}\,, \qquad \frac{dr}{d\rho} = -r \sqrt{f(r)}\,.
\end{equation}
with $\rho\geq 0$. In these coordinates the horizon is situated at $\rho=0$, while the conformal boundary is located at $\rho\to\infty$. This coordinate patch exclusively covers the region outside the black hole. However, as discussed in \cite{Caceres:2022smh}, the black hole's interior can be accessed by analytically continuing the radial coordinate, i.e., $\rho\to -i\rho$.\footnote{The time coordinate is shifted by a constant amount: $t=t_I-\text{sgn}(t_I)\gamma/2T$ where $\gamma$ is a half-integer and $T$ is the black hole temperature.}
In this patch, our thermal $a$-function $a^{\text{E}}_T(\rho)$ can be written as follows:
\begin{equation}\label{eq:domafunc}
  a^{\text{E}}_T(\rho) =\frac{\pi^{d/2}}{\Gamma \left(\frac{d}{2}\right)} \left(\frac{1}{\ell_P}\right)^{d-1}\bigg(\frac{h(\rho)}{A'(\rho)}\bigg)^{d-1}\exp\bigg[-\sum_{n=2}^{\infty}{\tilde{\lambda}_n}\int A'^{2({n-1})}\bigg(\frac{A''(\rho)h(\rho)-h'(\rho)A'(\rho)}{h(\rho)A'(\rho)}\bigg)d\rho\bigg].   
\end{equation}
To analyze the zero temperature case, we set $h=1$. In this case, the integrals in \eqref{eq:domafunc} can be explicitly carried out, leading to
\begin{eqnarray}\label{eq:domainwallafunc}
    a^{\text{E}}_T&=&\frac{\pi^{d/2}}{\Gamma \left(\frac{d}{2}\right)} \left(\frac{1}{\ell_P}\right)^{d-1}\bigg(\frac{1}{A'(\rho)}\bigg)^{d-1}\exp\bigg(-\sum_{n=2}^{\infty}\frac{\tilde{\lambda}_n}{2n-2}A'^{2({n-1})}\bigg)\,,\\
    &\approx&\frac{\pi^{d/2}}{\Gamma \left(\frac{d}{2}\right)} \left(\frac{1}{\ell_P}\right)^{d-1}\bigg(\frac{1}{A'(\rho)}\bigg)^{d-1}\bigg(1-\sum_{n=2}^{\infty}\frac{\tilde{\lambda}_n}{2n-2}A'^{2({n-1})}\bigg)\,,\label{eq:Myersafunc}
\end{eqnarray}
where in the last line we only kept the terms that are linear in $\tilde{\lambda}_n\ll 1$. Therefore, our $a$-function $a^{\text{E}}_T$ provides a (non-perturbative) generalization of the $a$-function introduced in \cite{Myers:2010jv} for a quasi-topological gravities with an arbitrary number of higher-derivative couplings.\footnote{Note that the $a$-function proposed in \cite{Myers:2010jv} and its generalization (\ref{eq:Myersafunc}) can become negative for sufficiently large couplings, indicating that they can only accurately represent a true $a$-function at the perturbative level.  In contrast, our $a$-function (\ref{eq:domainwallafunc}) remains positive definite for arbitrary values of the couplings.}

\subsection{Two-sided correlators}

In the AdS/CFT correspondence, there is a direct relationship between the length of boundary-anchored geodesics and two-point correlation functions of heavy operators in the boundary CFT \cite{Balasubramanian:1999zv,Louko:2000tp}. This entry in the holographic dictionary is crucial, as it illustrates how geometric data from the bulk encodes information about the boundary field theory.

More specifically, for two points \( x_1 \) and \( x_2 \) on the boundary of the AdS, the two-point correlation function \( G(x_1, x_2) \) of a primary operator \( \mathcal{O}(x) \) with large conformal dimension $\Delta\gg1$, is related to the bulk geodesic length \( L(x_1, x_2) \) according to:
\begin{equation}\label{twoptgeo}
G(x_1, x_2) \sim e^{-m \, L(x_1, x_2)}\,,
\end{equation}
where \( m\gg1 \) is the mass of the bulk scalar dual to \( \mathcal{O}(x) \). This expression is derived using a `first quantized' formalism for the correlator, followed by the application of a saddle-point approximation. The geodesic \( \gamma \) connecting the points \( x_1 \) and \( x_2 \) in the bulk minimizes the distance in the curved AdS space, ensuring it is the shortest path between these points. The length of this geodesic \( L(x_1, x_2) \), which extends deep into the bulk, thus provides a crucial map between geometric data of the AdS bulk and correlation functions in the dual CFT.

In the case of a two-sided black hole, it is well-established that space-like geodesics can traverse the Einstein-Rosen bridge, provided the boundary operators are inserted at the two disjoint AdS boundaries. In certain limits, these space-like geodesics become null, enabling them to approach arbitrarily close to the singularity.  This property was exploited in \cite{Fidkowski:2003nf,Festuccia:2005pi} to investigate how two-point CFT correlators encode information about the singularity,\footnote{More recently, similar calculations have been carried out in the presence shockwaves \cite{Horowitz:2023ury,Caceres:2023zft}, with the geodesic length in this case capturing information of certain higher-point correlators.} 
and was further revisited in \cite{Frenkel:2020ysx} in the context of holographic RG flows. Building on these pioneering works, we will now focus on calculating traversing geodesics to investigate potential signatures of the Kasner eons encoded in the boundary two-point correlators.

To simplify the calculations, we assume that the boundary operators are symmetrically placed, ensuring that the space-like geodesics anchored at these endpoints exhibit reflection symmetry. The radial geodesics originate from one of the AdS boundaries, travel towards the singularity until they reach a turning point, and then return to the second AdS boundary. Further, they are confined to the \( (t, r) \) plane, where the induced metric is given by,
\begin{equation}
ds^2 = \frac{1}{r^2} \left( -f(r)e^{-\chi(r)} dt^2 + \frac{dr^2}{f(r)} \right)\,.
\end{equation}
We can parametrize the geodesics using a single function, \( r = r(t) \), allowing us to express the length functional as:
\begin{equation}\label{laggeo}
L = \int dt \frac{1}{r} \sqrt{-f(r)e^{-\chi(r)} + \frac{\dot{r}^2}{f(r)}} \equiv \int dt \mathcal{L}_g\,,
\end{equation}
where \( \mathcal{L}_g \) represents the `Lagrangian' of the geodesic, defined implicitly by \eqref{laggeo}.

Since \( \mathcal{L}_g \) does not explicitly depend on time, we can define 
a conserved quantity \( E \), representing the energy associated with the space-like geodesic,
\begin{equation}\label{eq:conservE}
E = \dot{r} \frac{\partial \mathcal{L}_g}{\partial \dot{r}} - \mathcal{L}_g = \frac{f(r)e^{-\chi(r)}}{r \sqrt{-f(r)e^{-\chi(r)} + \frac{\dot{r}^2}{f(r)}}}\,.
\end{equation}
The geodesic length is then obtained by substituting \( E \) into \( \mathcal{L}_g \), yielding the minimal geodesic length anchored at some boundary time slice \( t = t_b(E) \),
\begin{equation}
L = \frac{2}{|E|} \int_{r_c}^{r_t} \frac{e^{-\chi/2} dr}{r^2 \sqrt{1 + \frac{f(r)e^{-\chi(r)}}{r^2 E^2}}} + 2 \log r_c\,.
\end{equation}
Note that a counterterm has been introduced to subtract the UV divergences arising from the AdS boundary. Here, \( r_c \) represents the UV cutoff, while \( r_t \) is the turning point of the geodesic, determined by
\begin{equation}
E^2 =g_{tt}(r_t)=\frac{-f(r_t)e^{-\chi(r_t)}}{r_t^2}\,.
\label{enturn}
\end{equation}
Using (\ref{eq:conservE}), we can express boundary time $t_b$ as a function of the energy $E$,

\begin{equation}
t_b = - P\int_0^{r_t} \frac{\text{sgn}(E)e^{\chi/2} dr}{f(r) \sqrt{1 + \frac{f(r)e^{-\chi(r)}}{r^2 E^2}}}\,,
\end{equation}
where $P$ denotes the principal value.
The relation in \eqref{enturn} specifies the depth of these geodesics within the black hole. Real space-like geodesics can be anchored up to a maximum boundary time $t_c$ given by
\begin{equation}\label{eq:tcrit}
    t_c=-P\int_{0}^{r_m}\frac{\text{sgn}(E)e^{\chi/2} dr}{f(r) \sqrt{1 + \frac{f(r)e^{-\chi(r)}}{r^2 E^2}}}\,,
\end{equation}
where $r_m$ is the maximum value of the turning point. 

\begin{figure}[t!]
\includegraphics[scale=0.60]{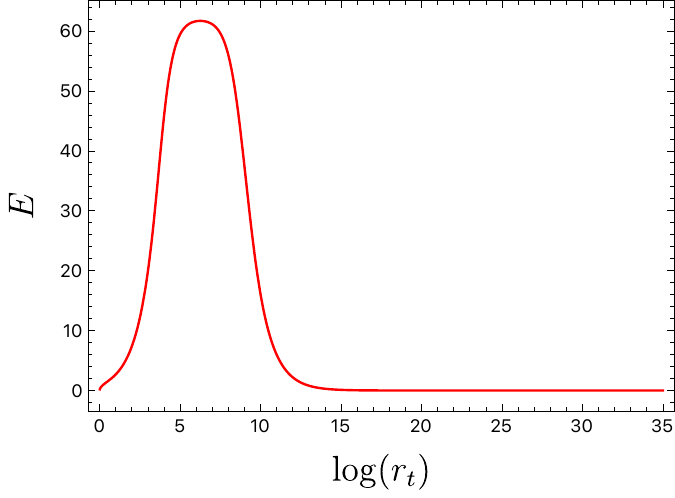}\hspace{10mm}
     \includegraphics[scale=0.61]{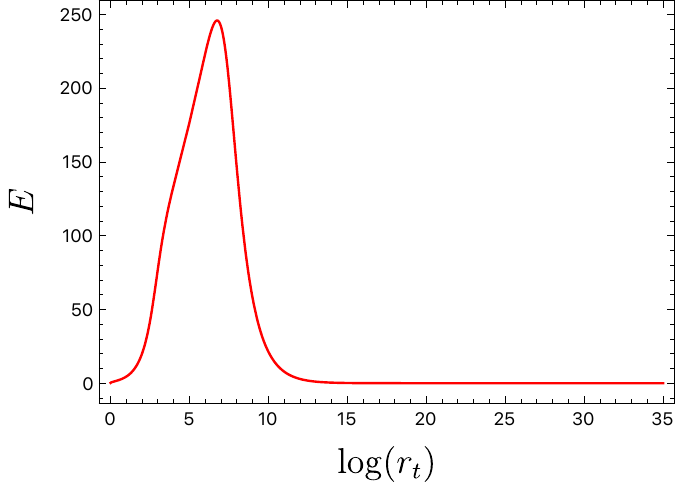}
     \caption{Energy of the two-sided geodesics that probe the black hole interior as a function of turning point $r_t$. We consider $d=4$ (left) and $d=5$ (right) with $\lambda_2=10^{-8}$, $\lambda_3=-10^{-22}$, $\lambda_4=-10^{-42}$, $\lambda_5=-10^{-67}$, $\lambda_{n>5}=0$ and $m^2=-\frac{d^2}{4}+\frac{1}{2}$. In both cases, the last eon has $p_t>0$ and the energy has therefore a maximum value. This implies that space-like geodesics get stuck at some finite $r$ and cannot reach arbitrarily close to the singularity.}
     \label{fig:energy_rt}
\end{figure}

\begin{figure}[t!]
\centering
\hspace{-8mm}
     \includegraphics[scale=0.67]{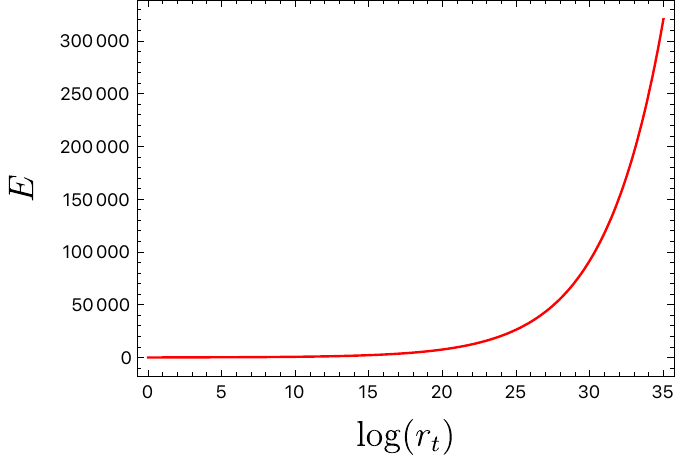}
     \includegraphics[scale=0.66]{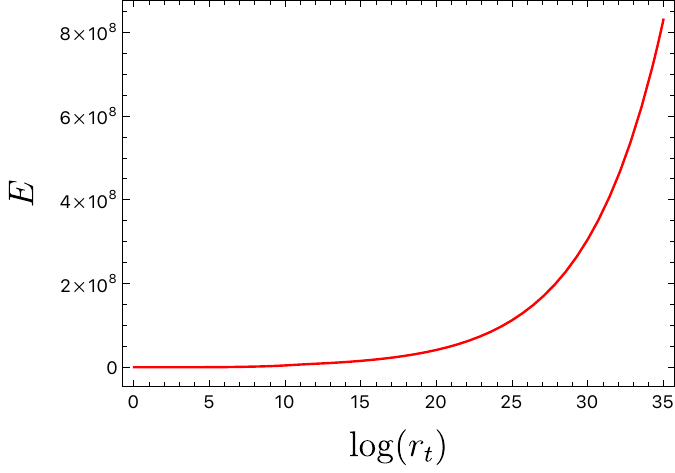}
     \caption{Left: energy of the two-sided correlator for Gauss-Bonnet theory without matter in $d=5$ with $\lambda_2=10^{-8}$ and $\lambda_{n>2}=0$. Right: the same energy evaluated for quasi-topological gravity theory up to quintic order without matter in $d=12$ with $\lambda_2=10^{-8}$, $\lambda_3=10^{-22}$, $\lambda_4=10^{-42}$, $\lambda_5=10^{-67}$ and $\lambda_{n>5}=0$. In both cases, the last eon has $p_t<0$, implying that space-like geodesics can reach arbitrarily close to the singularity.}
     \label{fig:energy_rt_GB}
\end{figure}

In Figures \ref{fig:energy_rt} and \ref{fig:energy_rt_GB}, we plot the energy of the geodesics as a function of the turning point for different scenarios of interest. For a two-sided geodesic to reach the singularity, its energy must diverge, i.e., $E\to\infty$. However, when the singularity resides within a Kasner eon characterized by $p_t>0$, there is always a finite maximum energy, hindering direct probing of the singularity. This phenomenon was first observed in \cite{Hartnoll:2020rwq}.\footnote{Charged RG flows, like those explored in \cite{Hartnoll:2020rwq}, typically result in $p_t>0$. It has recently been demonstrated that in these cases, charged geodesics can successfully probe the singularity \cite{Carballo:2024hem}. However, since the RG flows examined in this paper are neutral, the use of such geodesics is not feasible.} Notably, $p_t>0$ is quite generic for Kasner eons with matter (for $n\geq2$), as in these cases $p_t=n/d$, cf. \eqref{eq:kasnerexpscalarmet}. This behavior is depicted in Figure \ref{fig:energy_rt}. Conversely, for Kasner eons without matter, we have $p_t=2n/d-1$ as shown in \eqref{eq:kasnerexpeonvac}. Therefore, for $n\leq d/2$, we have $p_t\leq 0$, and space-like geodesics are expected to reach the singularity as $E\to\infty$.\footnote{For $n= d/2$ ($p_t=0$) $E$ has a finite maximum but the geodesic still reaches the singularity.} This behavior is illustrated in Figure \ref{fig:energy_rt_GB}.

\begin{figure}[t!]\hspace{-6mm}
\includegraphics[scale=0.58]{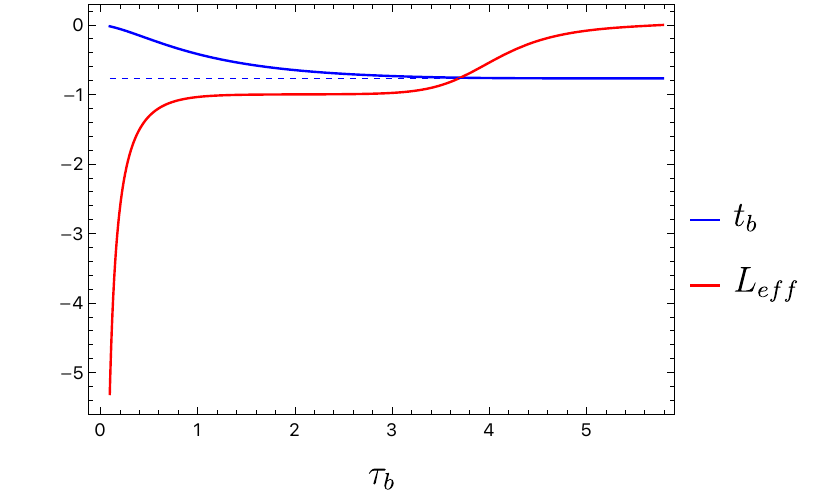}
     \includegraphics[scale=0.58]{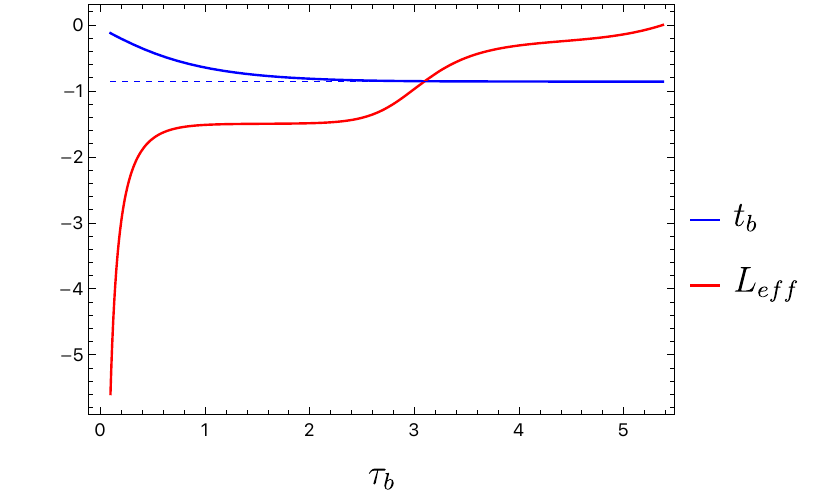}
\vspace{-6mm}
     \caption{$L_{\text{eff}}$ and $t_b$ as function of $\tau_b$ in $d=4$ (left) and $d=5$ (right) with $\lambda_2=10^{-8}$, $\lambda_3=-10^{-22}$, $\lambda_4=-10^{-42}$, $\lambda_5=-10^{-67}$, $\lambda_{n>5}=0$ and $m^2=-\frac{d^2}{4}+\frac{1}{2}$. The blue dashed line corresponds to the critical time, $t_c$. When this time is reached, the space-like geodesic reaches its maximum depth, $r_m$, situated in the last eon where $p_t\leq0$.}
     \label{fig:Leff_tb_rt}
\end{figure}

Whether or not the final eon has $p_t>0$, space-like geodesics can still probe all preceding eons with $p_t\leq0$. To effectively diagnose these eons, we find it convenient to perform a time parametrization, $t_b\to\tau_b(t_b)=\log (r_t(t_b))$, and define the following combination:
\begin{equation}\label{eq:Leff_def}
    L_{\text{eff}}(\tau_b)\equiv\frac{L''(\tau_b)}{L'(\tau_b)}-\frac{t''_b(\tau_b)}{t'_b(\tau_b)}=\frac{p_t^{\text{eff}}(\tau_b)}{p_x^{\text{eff}}(\tau_b)}\,,
\end{equation}
which perfectly captures the $p_t\leq0$ Kasner eons, as illustrated in Figure \ref{fig:Leff_tb_rt}. This provides a new means to identify stringy effects within the black hole interior. Crucially, $L_{\text{eff}}$ can be expressed entirely in terms of boundary quantities and their derivatives, with $L$ effectively encoding the two-point correlator as described in \eqref{twoptgeo}. However, a subtle but noteworhy caveat arises: \eqref{eq:Leff_def} presupposes a preferred notion of time, which may be difficult to identify from a purely boundary analysis (in part because the two-point correlators do not cease to exist at $t_b=t_c$, while the space-like geodesics do). An intriguing direction for future research would be to explore whether complex geodesics can reveal information about the correlator for $t_b>t_c$ and, in particular, if they can be used to diagnose the remaining eons with $p_t>0$. We leave this investigation for future work.

\subsection{Generalized complexity}

A further probe of the black hole interior that is worth investigating is quantum complexity \cite{Susskind:2014rva,Susskind:2014moa}. In \cite{Belin:2021bga, Belin:2022xmt}, the authors uncovered a broad class of gravitational observables exhibiting features typical of complexity, including late-time linear growth and the switchback effect. This framework, known as complexity=anything or generalized complexity, introduces novel geometric constructs that probe the bulk and have holographic connections to various notions of quantum complexity in the boundary theory.\footnote{The definition of quantum complexity in field theories is fraught with inherent ambiguities \cite{Nielsen:2006cea,Dowling:2006tnk,Jefferson:2017sdb,Chapman:2017rqy, Caceres:2019pgf}, which are believed to correspond to various formulations of holographic complexities. While significant work remains to clarify this mapping, recent work has reinterpreted generalized complexities \cite{Caceres:2023ziv} in terms of Lorentzian threads or `gatelines' \cite{Pedraza:2021mkh,Pedraza:2021fgp,Pedraza:2022dqi}, offering a more intuitive and microscopic interpretation.}

In this section, we aim to investigate whether holographic complexity can be utilized to identify the presence of Kasner eons. Notably, the original complexity-volume (CV) prescription \cite{Stanford:2014jda} cannot be used to probe the singularity, as extremal volume surfaces are typically trapped at a finite $r$. However, CV is merely one of many holographic proposals. The relationship between holographic complexities and the behavior near spacetime singularities has been explored in various contexts \cite{Bolognesi:2018ion,Caputa:2021pad,Bhattacharya:2021nqj,Caceres:2022smh,Caceres:2022hei,Katoch:2023dfh,Jorstad:2023kmq,Arean:2024pzo,Carballo:2024hem}. Moreover, it is known that certain subclasses of generalized complexities can directly probe the singularity \cite{Jorstad:2023kmq,Arean:2024pzo}. Thus, it is crucial to examine how the Kasner exponents relate to key features of complexity, such as linear growth. In this section, we intend to employ a class of generalized complexities that probe the singularity to clarify their relationship with the Kasner exponents emerging in our particular family of higher derivative theories.

\begin{figure}[t!]
\centering
   \includegraphics[scale=0.8]{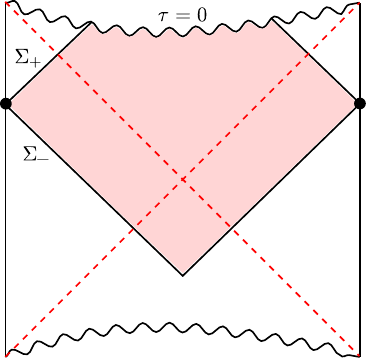} 
   \hspace{2cm}\includegraphics[scale=0.8]{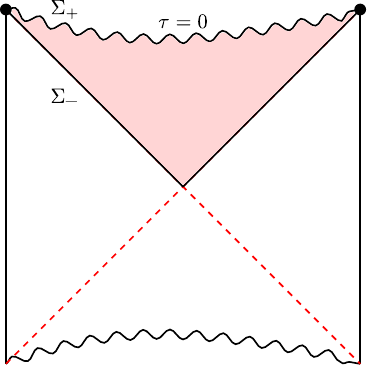} 
    \caption{Left: CMC slices $\Sigma_{\pm}$ in the limit of large extrinsic curvature, at a finite boundary time. The shaded region is the WdW patch. Right: the same situation but at late boundary times. In this case $\Sigma_+$ approaches the $\tau=0$ slice while $\Sigma_{-}$ approaches the horizon, $r=r_h$. }
    \label{CMCsliceatlatetimes}
\end{figure}

Let us start with a brief overview of \cite{Jorstad:2023kmq,Arean:2024pzo} and discuss how their approach can be adapted to diagnose the eons. The variants of complexity proposed in these studies define the complexity $\mathcal{C}$ in terms of a dimensionless scalar function, $\mathcal{F}$, evaluated on a co-dimension one hypersurface, $\Sigma$, characterized by constant mean curvature (CMC),
\begin{equation}
    \mathcal{C}=\frac{1}{G_N L}\int_{\Sigma}d^d y\sqrt{h}\, \mathcal{F}(y)\,,
\end{equation}
where $L$ is a length scale. The limit of particular interest occurs when the trace of the extrinsic curvature of the CMC slice diverges, $K\to\infty$. There are two such surfaces, corresponding to the future and past slices of the Wheeler-deWitt (WdW) patch. See Figure \ref{CMCsliceatlatetimes} for an illustration. At late boundary times, one of these surfaces approaches the singularity, while the other approaches the horizon. We will focus our attention on the former.
 
We will assume the following Kasner metric for the eons
\begin{align}
\begin{split}
\mathrm{d} s^2 =  -c_\tau^2 \, \mathrm{d}\tau^2 + c_t^2 \, \tau^{2 p_t} \mathrm{d} t^2      + \tau^{2 p_x} \mathrm{d} \vec x_{d-1}^{2}  \,.
\end{split}
\end{align}
This is equivalent to \eqref{eq:kasner}. However, 
we do not rescale $t$ and $r$ such that $c_r=c_t=1$ since the growth of complexity can depend on the choice of these coordinates, and these are naturally inherited from the RG flows. A quick calculation shows that the late-time rate of growth of complexity is given by: 
 \begin{equation}
     \frac{d\mathcal{C}}{dt}\bigg|_{t\to\infty}=\frac{c_tV_{d-1}}{G_N L}\,\mathcal{F}(\tau)\,\tau^{p_t+(d-1)p_x}\,,
 \end{equation}
where $V_{d-1}$ is the volume to the $(d-1)$-dimensional spaced spanned by $\vec{x}$. Reference \cite{Jorstad:2023kmq} considered $\mathcal{F}=L|K|$. Since $|K| = \frac{p_t+(d-1) p_x}{c_\tau \tau}$, using the Kasner relations of Einstein gravity one finds that the rate of growth is finite but independent of the Kasner exponents. A natural generalization for higher-curvature theories is $\mathcal{F}=L|K|^m$. In this case, we find that as $\tau\to0$,
\begin{equation}
    \frac{d \mathcal{C}}{dt}\bigg|_{t\to\infty}=\begin{cases}\,\,0 \quad\quad\quad\quad\quad\quad\quad\quad\quad\quad\quad\text{if}\quad 0\leq m<2n-1\,,\\ \frac{c_tV_{d-1}}{c_\tau^{2n-1} G_N}(2n-1)^{2n-1}\quad\quad\,\,\,\,\,\text{if}\quad m=2n-1\,,\\ \,\,\infty \quad\quad\quad\quad\quad\quad\quad\quad\quad\quad\,\,\,\text{if}\quad m>2n-1\,,\end{cases}
\end{equation}
for vacuum eons, while
\begin{equation}
    \frac{d \mathcal{C}}{dt}\bigg|_{t\to\infty}=\begin{cases}\,\,0 \quad\quad\quad\quad\quad\quad\text{if}\quad 0\leq m<n\,,\\ \frac{c_tV_{d-1}}{c_\tau^n G_N}n^n\quad\quad\,\,\,\,\,\,\,\text{if}\quad m=n\,,\\ \,\,\infty \quad\quad\quad\quad\quad\,\,\,\text{if}\quad m>n\,,\end{cases}
\end{equation}
for matter eons with a scalar field. To arrive at these expressions we have used the linear Kasner relations \eqref{eq:kasnerrelnew} and \eqref{eq:kasnerrelnewscalar}, respectively.

Reference \cite{Arean:2024pzo} considered an alternative formulation of complexity, where the functional $\mathcal{F}$ is constructed from another invariant derived from the extrinsic curvature tensor: $\mathcal{F}=L\sqrt{|K_{\mu\nu}K^{\mu\nu}|}$. By recognizing that $|K_{\mu\nu}K^{\mu\nu}|=\frac{p_t^2+(d-1)p_x^2}{c_\tau^2\tau^2}$, and applying the Kasner relations of Einstein gravity, one finds that the growth rate is finite and directly dependent on the Kasner exponents. This version of complexity, therefore, effectively diagnoses the nature of the singularity in Einstein gravity. A natural extension to higher-curvature theories is given by $\mathcal{F}=L|K_{\mu\nu}K^{\mu\nu}|^{m/2}$, where in the limit $\tau\to0$, we find,
\begin{equation}
\frac{d \mathcal{C}}{dt}\bigg|_{t\to\infty}=\begin{cases}\,\,0 \quad\quad\quad\quad\quad\quad\quad\quad\quad\quad\quad\quad\quad\,\,\,\,\text{if}\quad 0\leq m<2n-1\,,\\ \frac{c_tV_{d-1}}{c_\tau^{2n-1} G_N}\left(\frac{4n(n-1)}{d}+1\right)^{\frac{2n-1}{2}}\quad\quad\,\,\,\,\,\text{if}\quad m=2n-1\,,\\ \,\,\infty \quad\quad\quad\quad\quad\quad\quad\quad\quad\quad\quad\quad\,\,\,\,\,\,\,\text{if}\quad m>2n-1\,,\end{cases}
\end{equation}
for vacuum eons, while
\begin{equation}
\frac{d \mathcal{C}}{dt}\bigg|_{t\to\infty}=\begin{cases}\,\,0 \quad\quad\quad\quad\quad\quad\quad\quad\,\text{if}\quad 0\leq m<n\,,\\ \frac{c_tV_{d-1}}{c_\tau^n G_N}\left(\frac{n^2}{d}\right)^{\frac{n}{2}}\quad\quad\,\,\,\,\,\,\,\text{if}\quad m=n\,,\\ \,\,\infty \quad\quad\quad\quad\quad\quad\quad\,\,\,\,\text{if}\quad m>n\,,\end{cases}
\end{equation}
for matter eons with a scalar field. To arrive at these expressions we have used both the linear and quadratic Kasner relations given in \eqref{eq:kasnerrelnew} and \eqref{eq:kasnerrelnewscalar}, respectively.

A couple of comments are in order. First, in the above expressions, $n$ is meant to represent the final eon in which the singularity resides. For example, if the higher-curvature theory is truncated at the Gauss-Bonnet coupling, i.e. $\lambda_{n>2}=0$, then $n=2$.\footnote{When $n\to\infty$, our complexity probes require modification. In this regime, quantum gravity corrections become significant, rendering our classical analysis inadequate.} We will typically consider a situation where the last eon exhibits linear growth in complexity, as this is the expected behavior in the dual CFT.\footnote{More precisely, quantum complexity is expected to saturate at late times. However, our analysis is purely classical, and this saturation would only emerge when incorporating bulk quantum corrections. See, e.g., \cite{Iliesiu:2021ari}.} A crucial difference with Einstein gravity, however, is that for higher-curvature theories the Kasner exponents are completely determined in terms of $n$. Given that the two cases discussed above result in growth rates that explicitly depend on $n$, we can conclude that both function choices yield equivalent insights into the Kasner scaling within the final eon, making them equally effective probes of the singularity.
In principle, we should be able to extract information about the remaining eons and their transitions by examining the full time evolution of complexity and analyzing its $t$-scaling across different time scales. Since this is computationally more involved, we defer this analysis to future work.

\section{Discussion\label{sec:discussion}}
In general relativity, the BKL analysis offers a crucial framework for understanding spacetime dynamics near space-like singularities. However, a general framework for higher-curvature gravity theories is currently lacking. Our work takes a step in this direction by examining a specific class of higher-derivative theories of arbitrary order minimally coupled to a scalar field. In particular, we showed that the presence of matter significantly alters the near-singularity behavior, clearly distinguishing it from the vacuum case.

In the first part of our paper, we considered a family of quasi-topological gravities in vacuum.  By incorporating an infinite number of higher-curvature couplings, we explicitly demonstrated the emergence of an infinite sequence of Kasner eons as a space-like singularity is approached. Each eon is governed by emergent physics at energy scales dictated by higher-order curvature terms, with transitions occurring to successively higher-order eons as the singularity nears. This phenomenon arises independently of the cosmological constant, underscoring it as a fundamental characteristic of gravity. We further showed that these higher-order eons result in modified Kasner relations that differ from those established in Einstein gravity. Nevertheless, each higher-order eon can still be represented by a Kasner metric, with the exponents adhering to new constraints based on both the order of the eon and the dimension of spacetime.

Introducing a minimally coupled free massive scalar field into our model significantly altered the near-singularity analysis. We first observed that the dynamics near the singularity remained unaffected by both the mass of the scalar field and the cosmological constant. As in the vacuum case, an infinite sequence of Kasner eons emerged. However, the presence of the scalar field led to a dramatic change in the Kasner exponents. While the vacuum scenario exhibited two distinct Kasner exponents, $p_t$ and $p_x$, the inclusion of the scalar field caused these exponents to become equal for all higher-order eons. This indicates that the scalar field enhances the isotropy of the Kasner solution, rendering the temporal direction indistinguishable from the spatial directions. Importantly, we found that the full set of Kasner exponents, including $p_\phi$, are fully determined by the order of the eon and the spacetime dimensionality, in stark contrast with GR, where the exponents are not fully fixed. We attributed this difference to the absence of higher-derivative corrections in the matter sector.

In the final part of our paper, we focused on cases with a negative cosmological constant and explored the emergence of eons in the interior of black holes with holographic duals. We examined three key holographic observables ---the thermal $a$-function, two-point functions of heavy operators, and holographic complexity--- as potential probes for understanding the behavior of spacetime near singularities.  We first introduce non-perturbative generalizations of the thermal $a$-function applicable to hairy black holes in quasi-topological gravities of any curvature order. This generalization was achieved by employing two distinct conditions: the Null Energy Condition (NEC) and the Null Curvature Condition (NCC). While these conditions are equivalent in Einstein gravity, they generally differ in higher-curvature theories. Based on these conditions, we constructed two monotonic functions characterizing holographic RG flows in the presence of a relevant deformation that effectively capture \emph{all} Kasner eons within the black hole. This demonstrates that these monotonic functions serve as robust diagnostic tools for identifying stringy effects near black hole singularities.

Next, we analyzed two-point correlation functions of heavy operators, holographically dual to space-like geodesics anchored at the AdS boundary. Our study revealed that the presence of a scalar field, which induces a positive $p_t$, generally prevents the geodesics from probing regions near the singularity. In contrast, in the absence of a scalar field, the geodesics can reach the singularity if the theory satisfies $n \leq d/2$, resulting in $p_t\leq 0$. Regardless of the singularity's characteristics, we demonstrated that space-like geodesics can capture all initial eons with $p_t<0$, providing, at least partially, a valuable complementary probe of stringy effects in the black hole interior. In the future, it would be interesting to investigate whether complex geodesics can reveal additional aspects of the singularity, particularly their potential to diagnose eons that remain inaccessible through standard geodesics.

Lastly, we investigated the growth rate of complexity within specific variants of the complexity=anything proposal. Evaluating complexity on a slice of constant mean curvature close to the singularity, we demonstrated that the last eon is effectively captured and can be diagnosed through late-time linear growth of complexity. While we did not delve into this in detail, we anticipate that other eons can be similarly identified by examining different $t$-scalings in the full-time evolution of complexity, which we plan to explore in future work.

Our work represents a crucial first step towards developing a comprehensive BKL-like analysis in higher-curvature gravity, paving the way for future theoretical advancements. Several interesting directions merit exploration. First, it would be interesting to analyze the approach to the singularity in four-dimensional black holes supplemented with higher-curvature corrections. Indeed, our study with quasi-topological gravities applies for spacetime dimensions $D \geq 5$, leaving the four-dimensional case uncovered. To explore it, one may resort to the so-called generalized quasi-topological gravities \cite{Bueno:2016xff,Hennigar:2017ego,Ahmed:2017jod}, which exist in four dimensions but whose equations for static and plane-symmetric equations are nevertheless much more involved. Investigation in this area will be conducted in the future. Second, we have focused only on higher-derivative theories coupled to a minimally coupled massive scalar field. We found that even in this relatively simple scenario, the presence of matter alters the GR predictions for the Kasner exponents, without inducing chaotic behavior. A significant future direction would be to explore more complex matter content, potentially giving rise to `mixmaster' behavior. This would allow for a deeper understanding of the interplay between Kasner epochs, eras, and eons. Third, it would be interesting to investigate the effects of considering higher-curvature corrections in the matter sector, as well as non-minimal couplings. As explained above, we expect such generalizations to give more intricate dynamics and more freedom in the definition of Kasner exponents and Kasner relations associated to the different eons. Finally, while our focus was on space-like singularities within black holes, examining cosmological applications may be of interest too. For instance, we can consider holographic cosmologies from dynamical end-of-the-world (ETW) branes falling into a black hole \cite{Cooper:2018cmb, Antonini:2019qkt, Fallows:2022ioc}. These branes, whose worldvolume is a big-bang/big-crunch cosmology, end in the singularity, and thus are sensitive to the presence of eons. Thus, it is natural to ask what are the implications of our work for holographically cosmologies constructed in this way. We hope to return to some of these questions in the near future.

\noindent \noindent\section*{Acknowledgments}
We are grateful to Daniel Areán, Pablo Bueno, Pablo A. Cano, Javier Carballo, Robie Hennigar, Hyun-Sik Jeong, Hanzhi Jiang, Gerben Oling, Le-Chen Qu and Simon Ross for useful discussions and correspondence.  EC and \'AM thank the Instituto de F\'isica Te\'orica UAM-CSIC (IFT UAM-CSIC) for hospitality
during the early stages of this project. The work of EC is supported by the National Science Foundation
under grants No. PHY-2112725 and No. PHY–2210562. \'AM acknowledges support from the Istituto Nazionale di Fisica Nucleare through the Bando 23590 and the INFN Special Initiative \textit{String Theory and Fundamental Interactions}. AKP and JFP are supported by the ‘Atracci\'on de Talento’ program (Comunidad de Madrid) grant 2020-T1/TIC-20495, by the Spanish Research Agency via grants CEX2020-001007-S and PID2021-123017NB-I00, funded by MCIN/AEI/10.13039/501100011033, and by ERDF A way of making Europe.

\appendix

\section{Quasi-topological gravities with first-order vacuum equations for 
\emph{any} static and plane-symmetric background}\label{app:quasith}

We consider a static and plane-symmetric ansatz of the form (\ref{eq:bbans}). For spacetime dimensions $d+1=D \geq 5$, it turns out that there always exists a higher-curvature gravity $\mathcal{Z}_{(n)}$ which is purely constructed out of curvature invariants of $n$-th curvature order and whose equations of motion on top of \eqref{eq:bbans} contain no more than two derivatives --- as a matter of fact, equations are of first-order in derivatives for $f(r)$ and $\chi(r)$. Such theories correspond to special subclasses of quasi-topological gravities and below we present explicit instances of such theories up to order $n=5$ and arbitrary dimension. We denote by $W_{abcd}$ the Weyl curvature tensor and by $Z_{ab}=R_{ab}-\frac{1}{D}g_{ab}R$ the traceless part of the Ricci tensor.
\begin{align}
\mathcal{Z}_{(1)}&=R\,, \\
\mathcal{Z}_{(2)}&=R^2- \frac{4D(D-1)}{(D-2)^2} Z^{ab} Z_{ab}+\frac{D(D-1)}{(D-2)(D-3)} W^{abcd} W_{abcd}\,, \\
\mathcal{Z}_{(3)}&=R^3+\frac{2  D^2 (D-1)^2 (2 D-3) W\indices{^a^b_c_d}W\indices{^c^d_e_f}W\indices{^e^f_a_b}}{(D-2)(D-3)  (D ((D-9)
   D+26)-22)}+\frac{24 D^2(D-1)^2  W\indices{_a_c^b^d} Z^a_b Z^c_d}{(D-3) (D-2)^3}\notag\\
   &-\frac{24 D^2 (D-1)^2   W_{a c d e}W^{bcde}Z^a_b}{(D-3) (D-4)
   (D-2)^2}+\frac{16 D^2 (D-1)^2
   Z^a_b Z^b_cZ_a^c}{(D-2)^4}+\frac{3 (D-1) D R W_{a b c d} W^{a b c d}}{(D-2) (D-3)}\notag\\
   &-\frac{12 D (D-1) 
   Z_a^b Z^a_b R}{(D-2)^2}\, ,\\ \nonumber
\mathcal{Z}_{(4)}&=R^4+\frac{3 D^3(D-1)^2 (3 D-4) \left(W^{a b c d} W_{a b c d}\right)^2}{(D-2)^4
   (D-3)^2}- \frac{384 D^3 (D-1)^3  Z_{a c} Z_{d e} W^{bdce} Z^{a}_b}{(D-3)
   (D-4) (D-2)^4}\\\nonumber &+\frac{8 D^2(D-1)^2 (2
   D-3) R W\indices{^a^b_c_d}W\indices{^c^d_e_f}W\indices{^e^f_a_b}}{(D-2) (D-3) (D ((D-9)
   D+26)-22)}+\frac{48 D^3 (D-1)^3
  \left(Z_a^b Z^a_b\right)^2}{(D-2)^5 (D-3)}\\\nonumber & -\frac{96 D^3 (D-1)^3 
  Z_a^bZ_b^cZ_c^dZ_d^a}{(D-2)^5(D-3) } +\frac{64 D^2(D-1)^2 R
   Z_a^b Z_b^cZ^a_c}{(D-2)^4}+\frac{96 D^2 (D-1)^2 R  W\indices{_a_c^b^d}Z^a_b Z^c_d}{
   (D-2)^3 (D-3)}\\\nonumber & -\frac{96 D^2 (D-1)^2  R
   W_{a c d e}W^{bcde}  Z^a_b }{(D-3) (D-4) (D-2)^2}-\frac{48 D^2 (D-1)^4 W_{a b c d} W^{a b c d}Z_{ef} Z^{ef}}{(D-3)
   (D-2)^4}\\\nonumber &-\frac{192 D^2 (D-1)^3  W_{acbd} W^{c efg} W^d{}_{efg} Z^{ab}}{
   \left(D^2-5 D+6\right)^2 (D-4)}+\frac{6 D (D-1)  R^2
 W^{a b c d} W_{a b c d}}{(D-2) (D-3)}\\&-\frac{24 D(D-1)  R^2
   Z^{ab} Z_{ab}}{(D-2)^2}+\frac{96 D^2(2D-1)(D-1)^3 W_{abcd} W^{aecf} Z^{bc} Z_{ef}}{(D-2)^3(D-3)^2}\,,
\end{align}
\begin{align}
\nonumber
\mathcal{Z}_{(5)}&=R^5+\frac{4 (4 D-5) D^4(D-1)^3   W\indices{_a_b^c^d}W\indices{_c_d^e^f}W\indices{_e_f^a^b}  W^{ghij} W_{ghij}
 }{(D-3)^2 (D-2)^3 (D ((D-9)
   D+26)-22)}\\\nonumber &-\frac{640 D^4 (D-1)^4 Z^{de} Z_{de}
   Z_a^b Z_b^cZ^a_c}{(D-3) (D-4) (D-2)^6}+\frac{15 D^3 (D-1)^2
  (3 D-4) R \left (W^{abcd} W_{abcd} \right)^2}{(D-3)^2
   (D-2)^4}\\\nonumber &+\frac{48 D^4 (D-1)^3 \left(16 (D-1)
   (D-3) Z^{a}_b  Z^{b}_{c} Z^{c}_d Z^{d}_{f} Z^{f}_a-5 (D-2)^2 W^{ghij} W_{ghij}  W_{a c d e}W^{bcde} Z^a_b\right)}{(D-4)
   (D-3)^2 (D-2)^6}\\\nonumber &-\frac{1920 D^3 (D-1)^3  R W^{bdce} Z^{a}_b Z_{a c} Z_{d e} }{(D-3)
   (D-4) (D-2)^4}+\frac{240 D^3 (D-1)^3  R
   \left ( Z^{ab} Z_{ab} \right)^2}{(D-3) (D-2)^5}\\\nonumber &-\frac{480 D^3 (D-1)^3 R
  Z_a^bZ_b^cZ_c^dZ_d^a}{(D-3) (D-2)^5}-\frac{320 (D-1)^4 D^3
   (D(D+2)-2) W_{ghij} W^{ghij} Z^a_b Z^b_cZ_a^c}{(D-4) (D-3)
   (D-2)^6 (D+1)}\\\nonumber & +\frac{40 D^4 (D-1)^3  (9 D^4-88 D^3+307 D^2-440 D+206) W\indices{_a_b^c^d}W\indices{_c_d^e^f}W\indices{_e_f^a^b} Z_{gh} Z^{gh}}{ (D-2)^3 (D-3)(D-4) (D^2-6D+11) (D
   ((D-9) D+26)-22)}\\\nonumber & +\frac{480 D^4 (D-1)^4  Z^{ef} Z_{ef}
   W\indices{_a_c^b^d} Z^a_b Z^c_d }{(D-3)^2 (D-2)^5} +\frac{20 D^2 (D-1)^2  (2
   D-3) R^2 W\indices{_a_b^c^d}W\indices{_c_d^e^f}W\indices{_e_f^a^b}}{(D-2) (D-3) (D ((D-9)
   D+26)-22)}\\\nonumber & +\frac{240 D^2 (D-1)^2 R^2
   W\indices{_a_c^b^d} Z^a_b Z^c_d }{(D-3) (D-2)^3}+\frac{160 D^2 (D-1)^2 R^2
   Z_a^b Z_b^cZ^a_c}{(D-2)^4}\\\nonumber & -\frac{960 D^2 (D-1)^3  R  W_{acbd} W^{c efg} W^d{}_{efg} Z^{ab}}{(D-4)
   \left(D^2-5 D+6\right)^2} -\frac{240 D^2 (D-1)^2 R^2
    W_{a c d e}W^{bcde}Z^a_b}{(D-3) (D-4) (D-2)^2}\\\nonumber & -\frac{240 D^2 (D-1)^4
    R W_{abcd} W^{abcd} Z_{ef}Z^{ef}}{(D-3)
   (D-2)^4}+\frac{10 D (D-1) D R^3
   W^{abcd} W_{abcd}}{(D-2) (D-3)}\\&-\frac{40 D (D-1)  R^3
   Z^{ab} Z_{ab}}{(D-2)^2}+\frac{480 D^2(2D-1)(D-1)^3 R W_{abcd} W^{aecf} Z^{bc} Z_{ef}}{(D-2)^3(D-3)^2}\nonumber \\
   &-\frac{1920 D^3 (3D-1)(D-1)^4 Z^{ab} W_{acbd} Z^{ef}  W\indices{_e^c_f^g} Z^{dg}}{(D-2)^4 (D-3)^2 (D-4)(D+1)}-\frac{960 D^4 (D-1)^4 Z_{a}^b Z_{b}^{c} Z_{cd} Z_{ef} W^{eafd}}{(D-2)^5(D-3)^2}\nonumber \\&
   -\frac{240D^4 (D-1)^3(3D-2) Z^a_b Z^b_c W_{daef} W^{efgh} W_{gh}{}^{dc}}{(D(D-6)+11)(D-2)^3(D-3)(D-4)}\,.
\end{align}

Having at disposal quasi-topological gravities with first-order vacuum equations on any static and plane-symmetric background up to quintic order in curvature, it is possible to obtain such theories to arbitrary order in the curvature and any spacetime dimension $D$ via the following recursive formula, adapted from \cite{Bueno:2019ycr,Bueno:2024fzg}:
\begin{equation}
    \mathcal{Z}_{(n+5)}=\frac{3(n+3) R \, \mathcal{Z}_{(n+4)}}{n+1}-\frac{3(n+4)  \mathcal{Z}_{(2)} \mathcal{Z}_{(n+3)}}{n}+\frac{(n+3)(n+4) \mathcal{Z}_{(3)} \mathcal{Z}_{(n+2)}}{n(n+1)}\,,
    \label{eq:recrel}
\end{equation}
where $n \geq 1$. Surprisingly enough, written in this way the recursive relation \eqref{eq:recrel} is dimension-independent. 

\section{Holographic dictionary for bulk scalar fields}

\label{app:holdic}
In this appendix, we discuss some basic entries of the holographic dictionary for the theory at hand, given by (\ref{eq:lagholo}). By varying the action with respect to the metric $g_{\mu\nu}$ and scalar field $\phi$, we obtain following equations of motions,
\begin{eqnarray}
    \phi''+ \left(\frac{f'}{f} - \frac{d-1}{r} - \frac{\chi'}{2}\right)\phi' + \frac{\Delta(d-\Delta)}{r^2 f}\phi& =& 0\,,\label{holeom1}\\
    \chi' - \frac{2f'}{f} - \frac{1}{\sum_{n=1}^\infty \tilde{\lambda}_n f^{n-1}}\bigg(\frac{\Delta(d-\Delta)\phi^2}{(d-1)rf} + \frac{2d(d-1)}{rf} - \frac{2d}{r} \sum_{n=1}^{\infty}\frac{\tilde{\lambda}_n}{n}f^{n-1}\bigg)&=&0\,,\label{holeom2}\\
    \chi'-\frac{r (\phi')^2}{\sum_{n=1}^\infty \tilde{\lambda}_n f^{n-1}}&=& 0\,,\label{holeom3}
\end{eqnarray}
where $m$ is expressed in terms of the scaling dimension $\Delta$ of the operator $\mathcal{O}$ dual to $\phi$ \cite{Witten:1998qj,Gubser:1998bc}:
\begin{equation}
m^2 = \Delta (\Delta - d)\label{massdim}\,.
\end{equation}
From this relation \eqref{massdim}, we see that the dual operator is relevant only if $\Delta < d$, which occurs when $m^2 < 0$. Using the Breitenlohner-Freedman stability bound \cite{Breitenlohner:1982bm,Breitenlohner:1982jf}, we can further determine a lower bound for $m^2$, which yields
\begin{equation}
  -\frac{d^2}{4}\leq m^2<0 \,.
\end{equation}
Additionally, for holographic applications, we impose the unitarity bound for $\Delta$ in order to have a unitary theory,
\begin{equation}
    \Delta\geq \frac{d-2}{2}\,.
\end{equation}
In the standard quantization, this bound further restricts $m^2$, so that\footnote{Larger masses are attainable by working in the alternative quantization.}
\begin{equation}
    -\frac{d^2}{4}\leq m^2\leq1-d^2/4 \,.
\end{equation}
We will thus consider massive scalar fields within this range.

To solve the equations in (\ref{holeom1})-(\ref{holeom3}), we need appropriate boundary conditions for bulk fields.  As the fields approach the boundary at $r\to0$ with $\Delta\neq d/2$, they behave as \cite{Caceres:2021fuw}:
\begin{align}\label{eq:bcfields}
\phi(r) &\sim \phi_0 r^{d-\Delta}\,,\\ 
\chi(r) &\sim \frac{d-\Delta}{2(d-1)\phi_0^2 r^{2(d-\Delta)}}\,,\\
f(r)&\sim 1\,, 
\end{align}
where $\phi_0$ is the source of the operator $\mathcal{O}$ dual to $\phi$. Note that for $\Delta=d/2$, the above expressions fail, and need to be modified. For this special case, $\Delta=d/2$, it has been found that they instead behave as \cite{Caceres:2021fuw},
\begin{align}
\phi(r) &\sim \phi_0 r^{d/2}\log r\,,\label{logPhi}\\
\chi(r) &\sim \frac{\phi_0^2}{4d(d-1)}r^d (2+2d \log r+d^2 (\log r)^2)\,,\\
f(r)&\sim 1\,, 
\end{align}
On the other hand, near the singularity as $r\to\infty$, they generally diverge as
\begin{equation}
    \!f(r)=-f_n r^{\gamma_n}\,,\qquad \chi(r)= \chi_n \log r+\chi_0\,, \qquad \phi(r)=\begin{cases} \phi_1 \log r \quad\quad\,\, (n=1)\,, \\ \phi_n r^{\frac{(n-1)\gamma_n}{2}}  \quad  (n\geq2)\,,
    \end{cases}
    \label{eq:holKasnersing}
\end{equation}
where $\chi_n=\gamma_n=\frac{2d}{n}$ for $n\geq 2$, $f_n>0$ and $\chi_0$ are some arbitrary constants and 
\begin{equation}
    \chi_1=\frac{\phi_1^2}{(d-1)}\,,\qquad  \gamma_1=d+\frac{\phi_1^2}{2(d-1)}\,.
\end{equation}

\bibliographystyle{JHEP-2}
\bibliography{Gravities.bib}

\end{document}